\documentclass[twocolumn,showpacs,aps,prl,superscriptaddress]{revtex4}
\usepackage{graphicx}
\usepackage{dcolumn}
\usepackage{amsmath}
\usepackage{epsfig}
\usepackage{multirow}
\usepackage{subfigure}

\input pubboard/babarsym

\newcommand{\BABARPubYear}    {06}
\newcommand{\BABARPubNumber}  {13}

\newcommand{\SLACPubNumber} {11933}
\newcommand{\LANLNumber} {xxxxxxx}

\def\mes   {\ensuremath{m_{ES}}}
\def\de   {\ensuremath{\Delta E}}
\def\bb   {\ensuremath{B \overline{B} }}
\def\Bz      {\ensuremath{B^0}\xspace}
\def\Bzb     {\ensuremath{\Bbar^0}\xspace}
\def\BzBzb   {\ensuremath{\Bz {\kern -0.16em \Bzb}}\xspace}

\def\bds   {\ensuremath{\Bz \to \Dstarm\proton\antiproton\pi^+}\xspace}
\def\bd   {\ensuremath{\Bz \to \Dm\proton\antiproton\pi^+}\xspace}
\def\bdsz {\ensuremath{\Bz \to \Dstarzb\proton\antiproton}\xspace}
\def\bdz   {\ensuremath{\Bz \to \Dzb\proton\antiproton}\xspace}

\def\dkp  {\ensuremath{\overline{D}^0 \to K^+\pi^-}}

\def\dkppz  {\ensuremath{\overline{D}^0 \to K^+\pi^-\pi^0}}

\def\bdsshort   {\ensuremath{\Bz\to\Dstarm\proton\antiproton\pi^+}}
\def\bdshort   {\ensuremath{\Bz\to\Dm\proton\antiproton\pi^+}}
\def\bdszshort {\ensuremath{\Bz\to\Dstarzb\proton\antiproton}}
\def\bdzshort   {\ensuremath{\Bz\to\Dzb\proton\antiproton}\xspace}

\def\rar   {\ensuremath{\rightarrow}}

\def\Lambdabar {\kern 0.2em\overline{\kern -0.2em \Lambda}{}\xspace}
\def\fbar {\ensuremath{\kern 0.2em\overline{\kern -0.2em f}{}}\xspace}

\def\figurebox#1#2#3{%
    \def\arg{#3}%
    \ifx\arg\empty
    {\hfill\vbox{\hsize#2\hrule\hbox to #2{\vrule\hfill\vbox to #1{\hsize#2\vfill}\vrule}\hrule}\hfill}%
    \else
    {\hfill\epsfbox{#3}\hfill}%
    \fi}

\begin{document}

\preprint{\babar-PUB-\BABARPubYear/\BABARPubNumber}
\preprint{SLAC-PUB-\SLACPubNumber}

\begin{flushleft}
\babar-PUB-\BABARPubYear/\BABARPubNumber\\
SLAC-PUB-\SLACPubNumber\\
hep-ex/\LANLNumber\\[10mm]
\end{flushleft}

\title{{\large \bf  Measurements of the Decays  $\bdz$, $\bdsz$, $\bd$, and $\bds$}}

%
\author{B.~Aubert}
\author{R.~Barate}
\author{M.~Bona}
\author{D.~Boutigny}
\author{F.~Couderc}
\author{Y.~Karyotakis}
\author{J.~P.~Lees}
\author{V.~Poireau}
\author{V.~Tisserand}
\author{A.~Zghiche}
\affiliation{Laboratoire de Physique des Particules, F-74941 Annecy-le-Vieux, France }
% ...
\collaboration{The \babar\ Collaboration}
\noaffiliation

\date{\today}

\begin{abstract}
We present measurements of branching fractions of $B^0$ decays to multi-body final states
containing protons,  based on 232 million  $\Upsilon(4S)\to
B\overline{B}$ decays collected with the \babar~detector at the SLAC
\pep2 asymmetric-energy $B$ factory. We measure  the branching fractions 
${\cal B}(\bdz)=(1.13\pm0.06\pm0.08)\times 10^{-4}$, 
${\cal B}(\bdsz)=(1.01\pm0.10\pm0.09)\times 10^{-4}$, 
${\cal B}(\bd)=(3.38\pm0.14\pm0.29)\times 10^{-4}$, and 
${\cal B}(\bds)=(4.81\pm0.22\pm0.44)\times 10^{-4}$ 
where the first error is statistical and the second systematic. We
present a search for the charmed pentaquark state, $\Theta_c(3100)$
observed by H1 and put limits on the branching fraction ${\cal B}
(\Bz \to \Theta_c \antiproton\pi^+)\times{\cal B}(\Theta_c \to \Dstarm\proton)<14\times10^{-6}$ and ${\cal B}
(\Bz \to \Theta_c \antiproton\pi^+)\times{\cal B}(\Theta_c\to D^-\proton)<9\times10^{-6}$. Upon investigation
of the decay structure of the above four $B^{0}$ decay modes, we see
an enhancement at low $p\overline{p}$ mass and deviations from  phase-space
in the $\overline{D}\overline{p}$ and $\overline{D}p$ invariant mass spectra.
\end{abstract}


\maketitle

The observations  of the $\bds$\cite{chconj} and $\ensuremath{B^{0}
\to D^{*-} p \overline{n}}$ decays by CLEO~\cite{prl_86_2732}, and
the $\bdz$ and $\bdsz$ decays by Belle~\cite{prl_89_151802} suggest
the dominance of multi-body final states in decays of \B mesons into
baryons~\cite{cheng} compared to two-body decays. In this paper we present measurements of
the branching fractions for the following four decay modes: $\bdz$,
 $\bdsz$, $\bd$, and $\bds$.  The study of the modes  presented here can
 help clarify the dynamics of weak decays of \B mesons involving
 baryons~\cite{Hou}.

Since the branching fractions of multi-body decays are
large~\cite{dunietz}, it is natural to ask whether such final
states are actually the products of  intermediate two-body
channels. If this is the case, then these initial two-body decays could
involve  proton-antiproton bound
states~($p\overline{p}$)~\cite{yang, rosner}, or  charmed pentaquarks~\cite{jaffe, wu}, 
or   heavy charmed baryons. Motivated by these
considerations, in particular the claim of a charmed pentaquark at
3.1 \gevcc by the H1 collaboration\cite{H1_collab}, the invariant mass
spectrum of the proton-antiproton and the invariant mass spectra of
the charmed meson and proton are investigated.  Throughout this
paper, we shall use the terms ``exotic'' and ``non-exotic'' to refer
to the ``$Dp$'' pair with total quark content $\overline{c}quud$ and
$\overline{c}q\overline{u}\overline{u}\overline{d}$ respectively (where $q$ is $u$ or $d$).
 Specifically, the ``exotic''
combinations refer to $D^{(*)-}p$ and $\overline{D}^{(*)0}p$ while the
``non-exotic'' combinations are $D^{(*)-}\overline{p}$ and $\overline{D}^{(*)0}\overline{p}$.

The data used in this analysis were accumulated with the \babar~detector~\cite{bib:babarNim} 
at the \pep2 asymmetric-energy $e^+e^-$
storage ring at SLAC. The data sample consists of  an integrated
luminosity of $212\pm2\invfb$  collected at the \FourS resonance
 corresponding to $(232\pm3)\times10^{6}$
$\B\Bbar$ pairs.  
The \babar~detector consists of 
a silicon vertex tracker (SVT) and a drift chamber (DCH) used for track and vertex
reconstruction,  an electromagnetic
calorimeter (EMC) for detecting photons and electrons, a Cherenkov detector (DIRC) and an instrumented flux return (IFR) used for particle identification (PID).  The efficiency of the selection criteria is determined with large samples of {\tt GEANT}-based \cite{geant} Monte Carlo (MC) simulated signal decays.  

We select $\overline{D}^0$ decays to $K^+\pi^-$,
$K^+\pi^-\pi^0$, and $K^+\pi^-\pi^+\pi^-$ and $D^-$
decays to $K^+\pi^-\pi^-$. We select $\overline{D}^{*0}$
decays to $\overline{D}^0\pi^0$ and $D^{*-}$ decays to $\overline{D}^0\pi^-$. The $B$
candidates are reconstructed from $D$ or $D^{*}$ candidates
combined with a proton and an antiproton track and a pion track if
appropriate. 
The $D$ candidates are required to have a mass within $\pm3\sigma$  of the  $D$ meson mass, $\widehat{m_D}$ \cite{pdg}.
The mass resolution, $\sigma(m_{D})$, ranges from 5.1 to 13.0 \mevcc for different $D$ decay channels,
the worst resolution corresponding to the mode with a $\pi^0$ in the final state. 
The $D^{*}$ candidates are selected by requiring  the mass difference $\Delta M= (m_{D\pi}- m_{D})$ to be within   $3\sigma$ of the nominal value, $\widehat{\Delta M}$,  where $\sigma \sim$ 1.0 \mevcc.  Particle identification  is required on the proton, antiproton, and
pion from the $B$, and on the kaon from the $D$ decay, using combined information from the 
energy loss, \dedx, in the SVT and the DCH and the Cherenkov angle in the DIRC. The proton 
identification efficiency is roughly 90\% with a mis-identification rate of less than 2\%.  
To suppress backgrounds of all kinds, vertexing probability
requirements are imposed on the $D$ and $B$ candidates.  In order to reduce background from
$e^+e^-\to q\overline{q}$ events (where $q$ is a $u$,$d$,$s$, or $c$ quark), the cosine of the angle
between the thrust axis of the $B$ candidate and that of the rest of the event
$|\cos{(\theta_{BT})}|$ is required to be less than 0.9
 and  the ratio of the second to the zeroth
Fox-Wolfram moments \cite{prl_41_1581} is required to be less than 0.35. 

 We select events in the region 5.2 \gevcc$<m_{ES}<$ 5.3 \gevcc and $|\Delta E|<$0.1 \gev, where $m_{ES} = \sqrt{(s/2+\textbf{p}_{\Upsilon}\cdot\textbf{p}_{B})^{2}/E_{\Upsilon}^{2}-\textbf{p}_{B}^{2}}$ ($\sqrt{s}$ is the total center-of-mass energy, $\textbf{p}_{B}$ is the $B$ meson momentum and ($E_{\Upsilon}$, $\textbf{p}_{\Upsilon}$) is the $\Upsilon(4S)$ 4-momentum,  defined in the laboratory frame), while $\Delta E = p_{\Upsilon}\cdot p_{B}/\sqrt{s}-\sqrt{s}/2$ ($p_{\Upsilon}$ = ($E_{\Upsilon}$, $\textbf{p}_{\Upsilon}$), $p_{B}$ = ($E_{B}$, $\textbf{p}_{B}$)).  The selection is kept loose because these two variables are used in a maximum likelihood fit  to extract the signal and background yields simultaneously.  
If there is more than one $B$ candidate passing these criteria for an event, the candidate is chosen that minimizes  $\chi^{2} = (m_{D}-\widehat{m}_{D})^2/\sigma(m_{D})^2 + (\Delta M - \Delta \widehat{M})^2/\sigma(\Delta M)^2$ for the modes $\bdsz$ and $\bds$,  and the candidate that minimizes  $\chi^{2} = (m_{D}-\widehat{m}_{D})^2/\sigma(m_{D})^2$
for the modes $\bdz$ and $\bd$.

The background for these modes comes from  $e^+e^-\rightarrow q\overline{q}$ events 
 and from $B$ decays other than those under consideration.
 In both of these cases, the background comes from selecting 
random combinations of tracks and thus does not peak in either $\Delta E$ or $\mes$.  
The one exception is in the case of $\bdsz$, where there is a possibility of events such
as  $B^0\rightarrow \overline{D}^0 p\overline{p}\pi^0$ 
that peak at the $B$ mass in $\mes$.  However, since the $\pi^0$ comes from the other B decay 
 in the event, the $\Delta E$ distribution does not peak strongly in the signal region.  

\begin{figure}[b]
\caption{ Fit projections of $\mes$  for (clockwise from top-left) $\bdz~(\overline{D}^0\to K^+\pi^-$), $\bdsz~(\overline{D}^0\to K^+\pi^-$), $\bds~(\overline{D}^0\to K^+\pi^-$), and $\bd~(\Dm\to \Kp\pi^-\pi^-)$. The dashed line is the  background contribution and the solid line is the background plus signal. }
\label{fig:Proj_B2}
 \centerline{   \epsfxsize4.5cm\epsffile{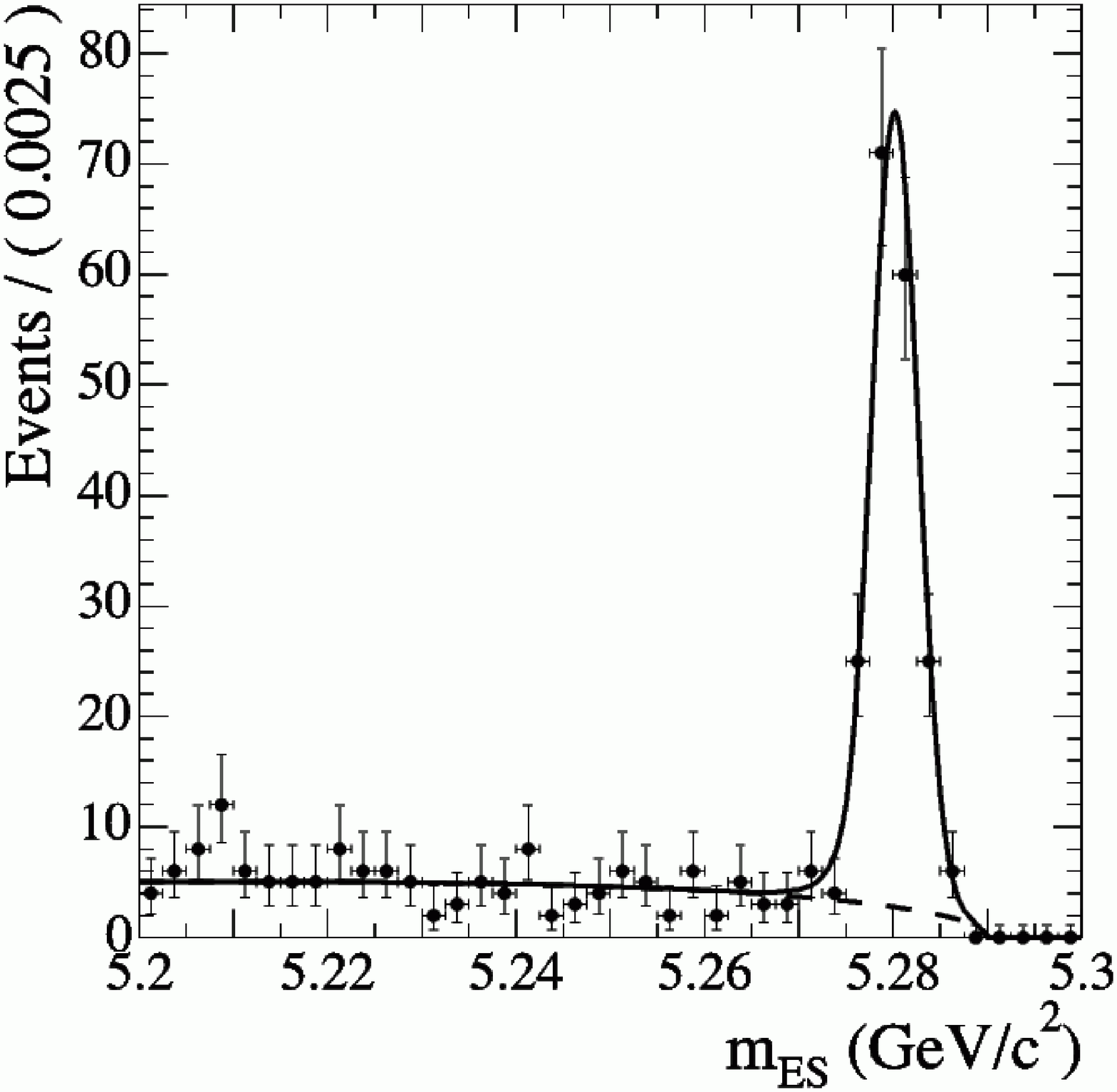}
                \epsfxsize4.5cm\epsffile{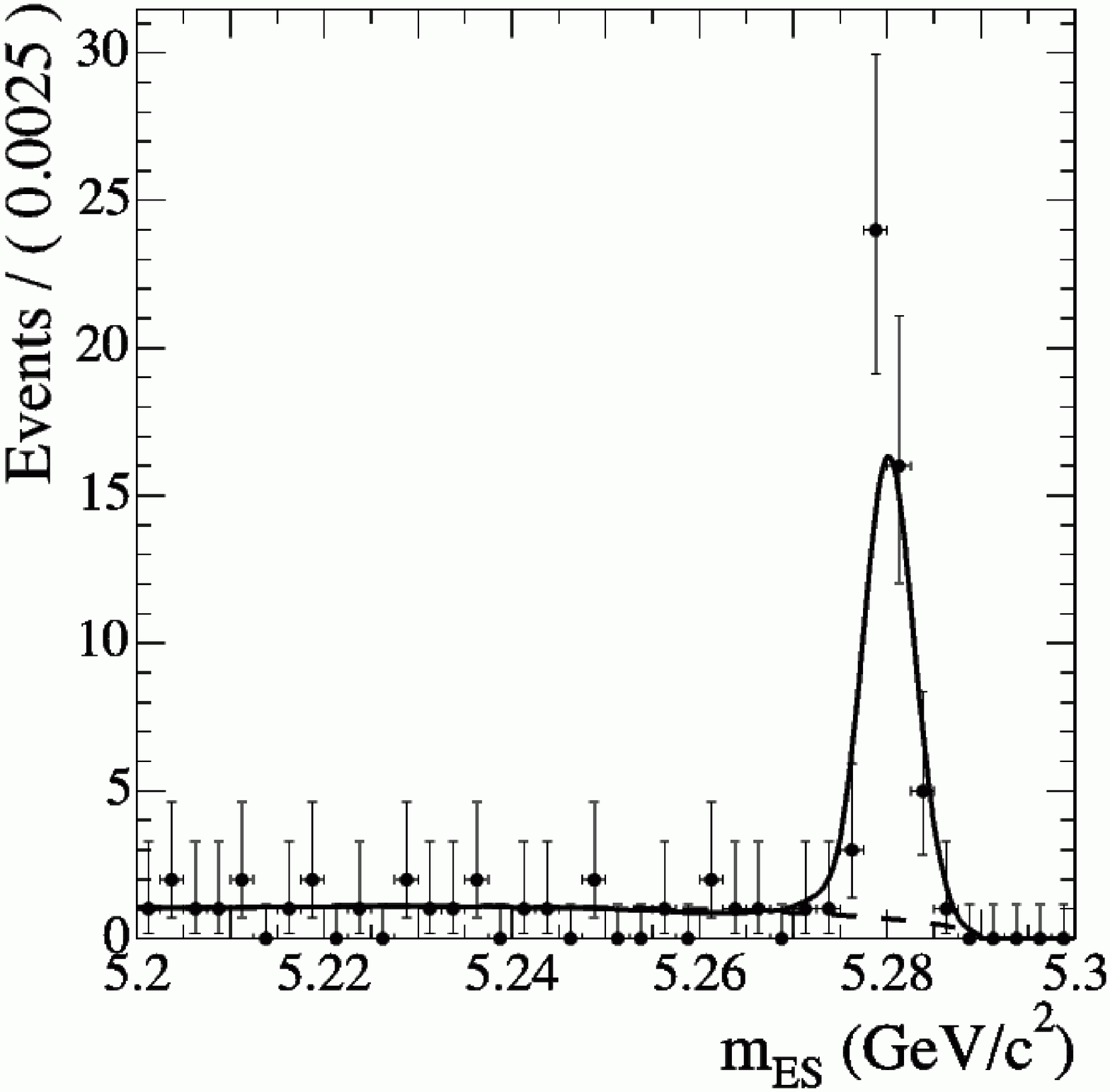}}
 \centerline{   \epsfxsize4.5cm\epsffile{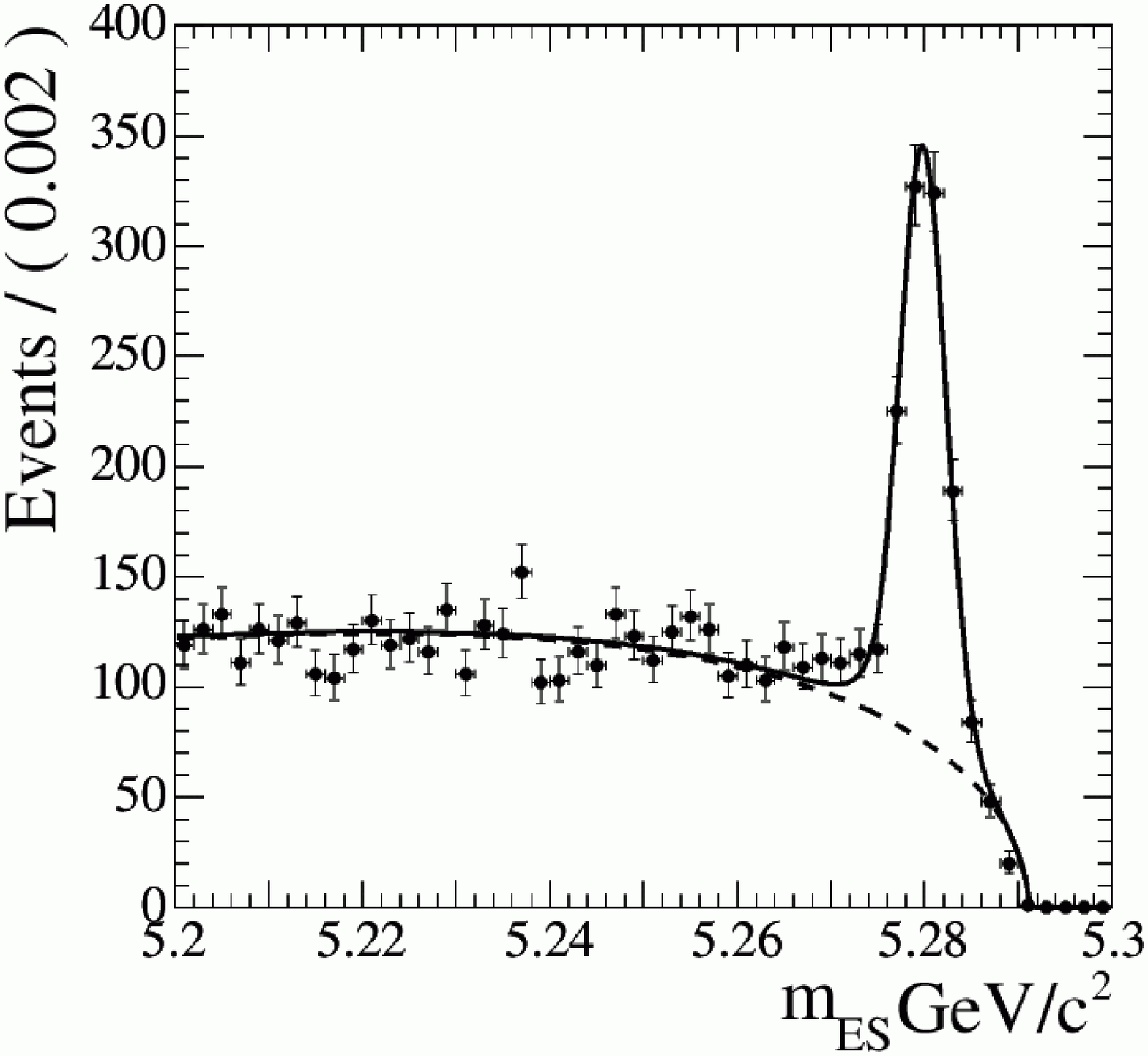}
                \epsfxsize4.5cm\epsffile{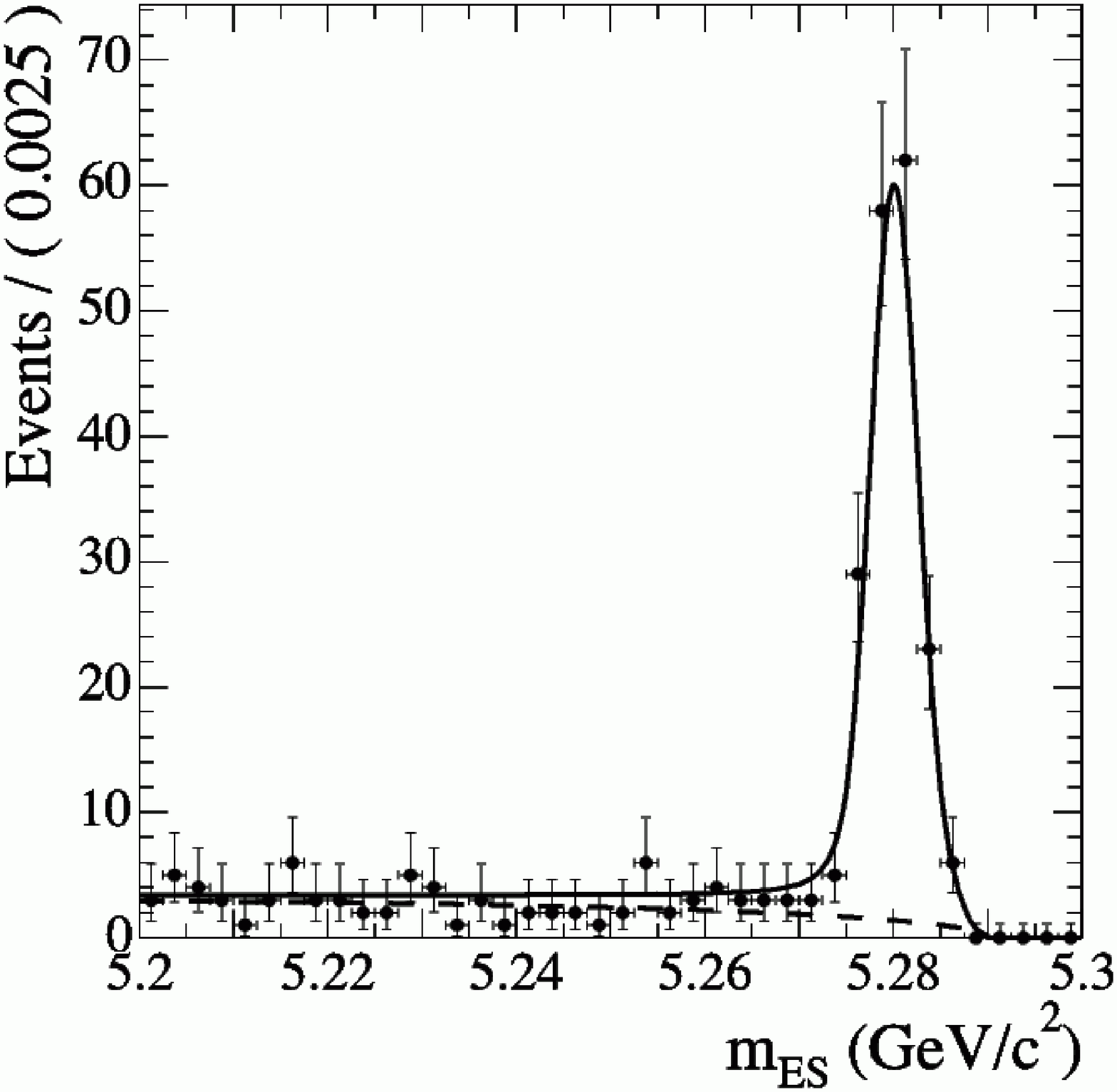}}
\end{figure}

We perform an unbinned extended maximum likelihood  fit to extract the yields.  The variables $\mes$ and $\de$ are used as discriminating variables to separate signal from background. The data sample is assumed to consist of two components: signal events and combinatorial background events due to random combinations of tracks from both $q\overline{q}$ and $\bb$ events.  For the decay $\bdszshort$, a peaking component is added to account for $B^0\rightarrow  \overline{D}^0 p\overline{p}\pi^0$ events.

In addition, the signal is split into correctly reconstructed events (Class I) and mis-reconstructed events (Class II).  The Class II events are signal events where one 
or more of the tracks from the signal B decay is lost and a track from the other $B$ decay is included in the reconstruction.
The fraction of Class II events is determined from MC and varies from nearly 0 for $\bdz\to\Kp\pi^- p\overline{p}$ to almost 50\% for $\bds\to\pi^-\Kp\pi^-\pi^0 p\overline{p}\pi^+$.  

In the maximum likelihood fit, each component is modeled by a probability density function (PDF) of the two variables $\mes$ and $\de$,
\begin{equation}
 {\cal P} = {\cal P}(\mes, \de).
\label{eqn:pdf}
\end{equation}
The  likelihood for the $N$ candidates in the event sample is given by:
\begin{equation}
\mathcal{L}=e^{-N^{\prime}}\cdot\prod_{i=1}^{N}\lbrace
N_{\rm sig}\cdot\lbrack f_{I} \cdot {\cal P}^i_{I}
+ f_{II}\cdot {\cal P}^i_{II} \rbrack
+N_{\rm bkg}\cdot {\cal P}^i_{\rm bkg}\rbrace,
\label{eqn:likelihood}
\end{equation}
where $N^{\prime}$ is the sum of the fitted number of signal ($N_{\rm sig}$) and background ($N_{\rm bkg}$) events. The background PDF is given by  ${\cal P}_{\rm bkg}$, ${\cal P}_{I}$ and ${\cal P}_{II}$ are the PDFs of Class I and II events in signal respectively, and $f_{I}$ and $f_{II}$ are their corresponding fractions.  

The Class I signal events are parameterized with a double
Gaussian for both $\mes$ and $\de$.  For Class II events, $\mes$ is
parameterized with the correlated function $P_{II}(\mes,
\de)=G(\mes)G_1(\de)+P(\mes)G_2(\de)$ where $G$ represents a Gaussian
and $P$ a polynomial function. All parameters for the signal PDFs
are obtained from signal MC and fixed in the fit with the exceptions
of the means of the narrow components of the double-Gaussian distributions for 
both $\mes$ and $\de$ for Class I events, which are
allowed to vary. The combinatorial background is parameterized with a
threshold function\cite{argus} in $\mes$ and a second-order
polynomial in $\de$, and all of the parameters are varied in the
fit.  The peaking  background component coming from $B$ decays in the $\bdsz$ modes is
modeled with a non-parametric 2-dimensional PDF in $\mes$ and
$\de$ and the yield is free in the fit. The $\mes$ distributions for the data and the fit, after selecting events with $|\de|<20\mev$, are shown in Figure~\ref{fig:Proj_B2} for the $\overline{D}^0\to K^+\pi^-$ and $\Dm\to \Kp\pi^-\pi^-$ decays.

For each event a signal weight is defined as follows: 
\begin{equation}
 W^i_{\rm sig} =
  \frac{ \sigma^2_{\rm sig} {\cal P}_{\rm sig}^{i}
   + \rm cov(\rm sig,\rm bkg) {\cal P}_{\rm bkg}^{i}}
   {N_{\rm sig}{\cal P}_{\rm sig}^{i} +
    N_{\rm bkg} {\cal P}_{\rm bkg}^{i}}, 
\label{eqn:br4}
\end{equation}
following the method described in Reference \cite{splots}. In Equation \ref{eqn:br4}, 
${\cal P}_{\rm sig}^{i}$ (${\cal P}_{\rm bkg}^{i}$) is the value of the signal (background) PDF for event $i$;  $\sigma_{\rm sig}$ is the standard deviation of the signal yield; and $\rm cov(\rm sig,\rm bkg)$ denotes 
the covariance between $N_{\rm sig}$ and $N_{\rm bkg}$, as obtained from the fit.  The normalization
 of $W^i_{\rm sig}$ is such that their sum equals the total number of signal events, $N_{\rm sig}$. 
 The sum of $W^i_{\rm sig}$ over a small area of phase space gives the correct distribution of signal in that area.

The branching fraction is obtained as:
\begin{equation}
 {\cal B} = \sum_{i} \frac{W^i_{\rm sig}}{N_{\bb} \cdot
                     \epsilon_i
                     \cdot {\cal B}_{\rm sub}},
\label{eqn:br3}
\end{equation}
where the sum is  over all events $i$, $N_{\bb}$ is the number of $\bb$ pairs in the sample, 
 $\epsilon_i$ is the efficiency for event $i$, which depends on its position in phase space,  
 and ${\cal B}_{\rm sub}$ is the product of the branching
fractions of the charmed meson decays \cite{pdg,cleoD}. 
We assume that the  $\Upsilon(4S)$ decays with equal probability to 
$\BzBzb$ and $B^+B^-$. 
The statistical error on the branching fraction is obtained from the fractional
error on the signal yield as calculated from the fit.  

The largest source of systematic error arises from the uncertainty in the charged track reconstruction efficiency determined from the MC.
This systematic error ranges from 3.3\% to 8.8\% depending on the number of charged tracks in the decay mode. 
In addition there is a systematic error due to the modeling of the PID efficiency for the protons and kaons of 4.5\% for all
modes and an additional error of 2\% for the pion identification for the modes $\bd$ and $\bds$.  
The uncertainty due to ignoring correlations between $\mes$ and $\de$ is estimated to be a few percent by performing fits to  Monte
Carlo samples that consist of fully simulated signal events
embedded with parameterized background events.  The
uncertainties related to modeling of the signal PDFs are calculated by allowing the $\de$ and $\mes$ signal shape parameters
for the $\bd$ mode to vary in the fit and then varying the fixed parameters in the other modes by the differences observed
between data and MC in this mode.  This error ranges from 0.2\% to 2.8\%.  
The fraction of Class II events is varied by $5\%$ per $\pi^0$, or $5\%$ for modes with no  $\pi^0$,
to account for the uncertainty due to mis-reconstructed events and the difference observed is 1\% to 5\%.
The uncertainty arising from binning the efficiency in  phase space gives a typical error of 3\%.
Finally, the errors on the branching fractions of $D$ and $D^*$ decays  are
included in the systematic uncertainty and range from 2.4\% (for $\bdz, \dkp$) to 6.2\% (for $\bdsz, \dkppz$).
The total systematic error ranges from 6.3\% to 13.3\%.  

\begin{table}[b]
\begin{center}
\caption{ The branching fractions (in units of $10^{-4})$ for the $\Bz$ decays considered here. The first
error is statistical and the second systematic.}
\label{tab:branching}
\begin{tabular}{l|l|c|c} \hline \hline
 $\Bz$ decay            & $D$ decay  &  $N_{\rm sig}$   & ${\cal B}(10^{-4})$ \\ \hline
                        & $K^+\pi^-$       &  214$\pm$16 &  \small{1.09$\pm$0.08$\pm$0.08} \\
$\bdzshort,$            & $K^+\pi^-\pi^0$  &  514$\pm$38 &  \small{1.15$\pm$0.08$\pm$0.10} \\ 
                        & $ K^+\pi^-\pi^+\pi^-$      &  320$\pm$26 &  \small{1.24$\pm$0.10$\pm$0.11}  \\\hline 

                        &  $K^+\pi^-$       & 57$\pm$9   &  \small{1.21$\pm$0.17$\pm$0.11} \\
$\bdszshort$            &  $K^+\pi^-\pi^0$ & 104$\pm$19  &  \small{1.08$\pm$0.14$\pm$0.14} \\ 
$\Dstarzb\to\Dzb\pi^{0}$&  $ K^+\pi^-\pi^+\pi^-$    & 46$\pm$12   &  \small{0.75$\pm$0.18$\pm$0.09}  \\\hline

$\bdshort$              & $ K^+\pi^-\pi^-$     & 1166$\pm$47 &  \small{3.38$\pm$0.14$\pm$0.29} \\ \hline

                        & $K^+\pi^-$        & 241$\pm$18  &  \small{4.84$\pm$0.40$\pm$0.44}\\
$\bdsshort,$            & $K^+\pi^-\pi^0$  & 522$\pm$32  &  \small{4.71$\pm$0.30$\pm$0.50} \\ 
$\Dstarm\to\Dzb\pi^{-}$ & $ K^+\pi^-\pi^+\pi^-$      & 311$\pm$24  &  \small{5.05$\pm$0.42$\pm$0.59} \\\hline\hline
\end{tabular}
\end{center}
\end{table}

The fitted signal yield and the measured branching fraction for each decay mode is given in Table \ref{tab:branching}.
Averaging the branching fractions of the different $D$ decays weighted by their errors and accounting for correlations, we obtain:
\begin{eqnarray*}
{\cal B}(\bdzshort)&=&(1.13\pm0.06\pm0.08)\times 10^{-4}\\
{\cal B}(\bdszshort)&=&(1.01\pm0.10\pm0.09)\times 10^{-4}\\
{\cal B}(\bdshort)&=&(3.38\pm0.14\pm0.29)\times 10^{-4}\\
{\cal B}(\Bz\to\Dstarm\proton\antiproton\pi^+)&=&(4.81\pm0.22\pm0.44)\times 10^{-4}
\end{eqnarray*}
where the first error is statistical and the second systematic.

\begin{figure*}
 \caption{The distributions of the  branching fractions (in units of $10^{-6}/\gevcc$) for $\bdz$ (top row, left), $\bdsz$ (top row, right), $\bd$ (middle row), and $\bds$ (bottom row) projected over two invariant mass dimensions.  
\label{fig:dalitz}}
\centerline{\epsfxsize5.cm\epsffile{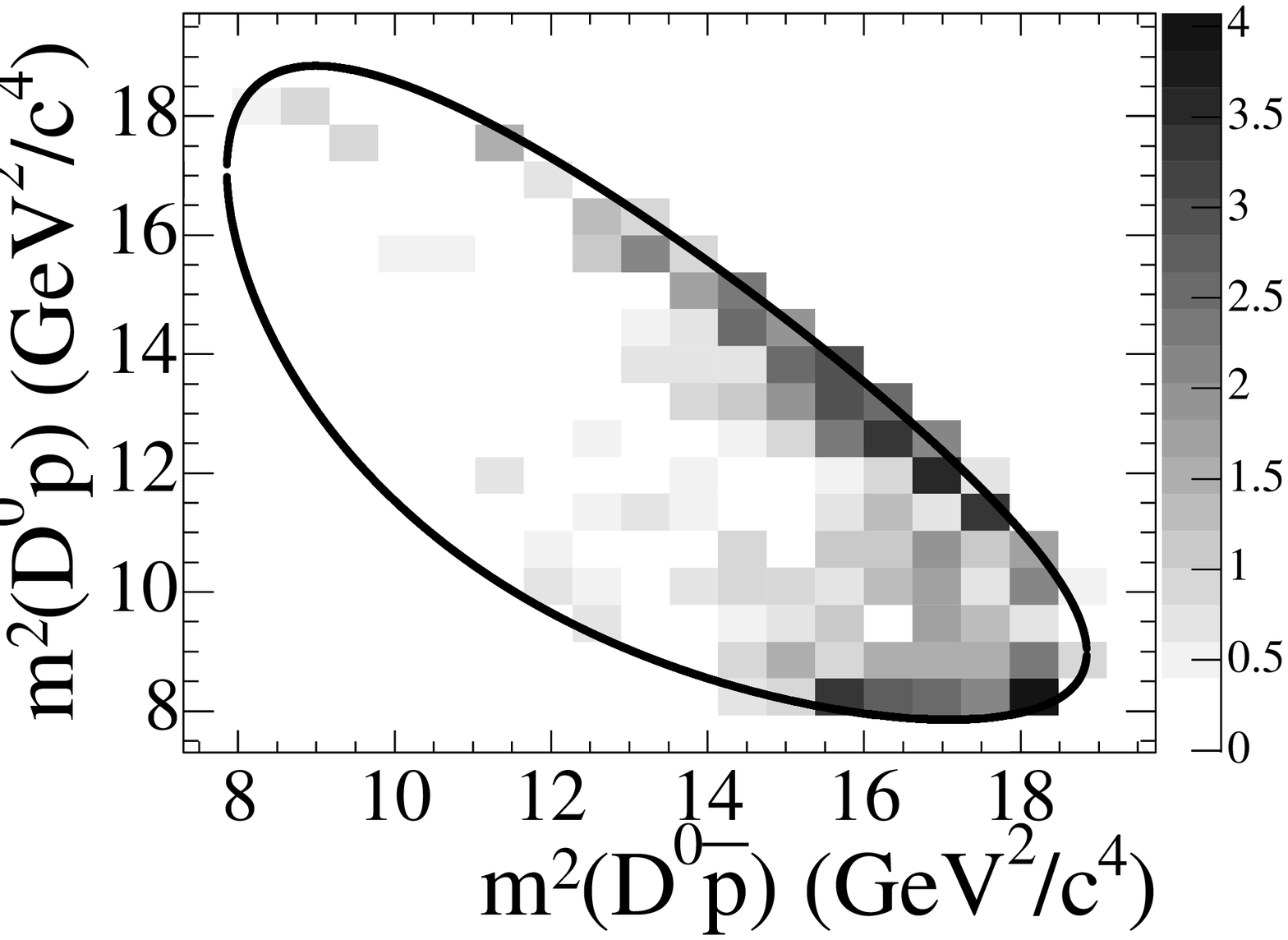}\hspace{0.1cm}
    \epsfxsize5.cm\epsffile{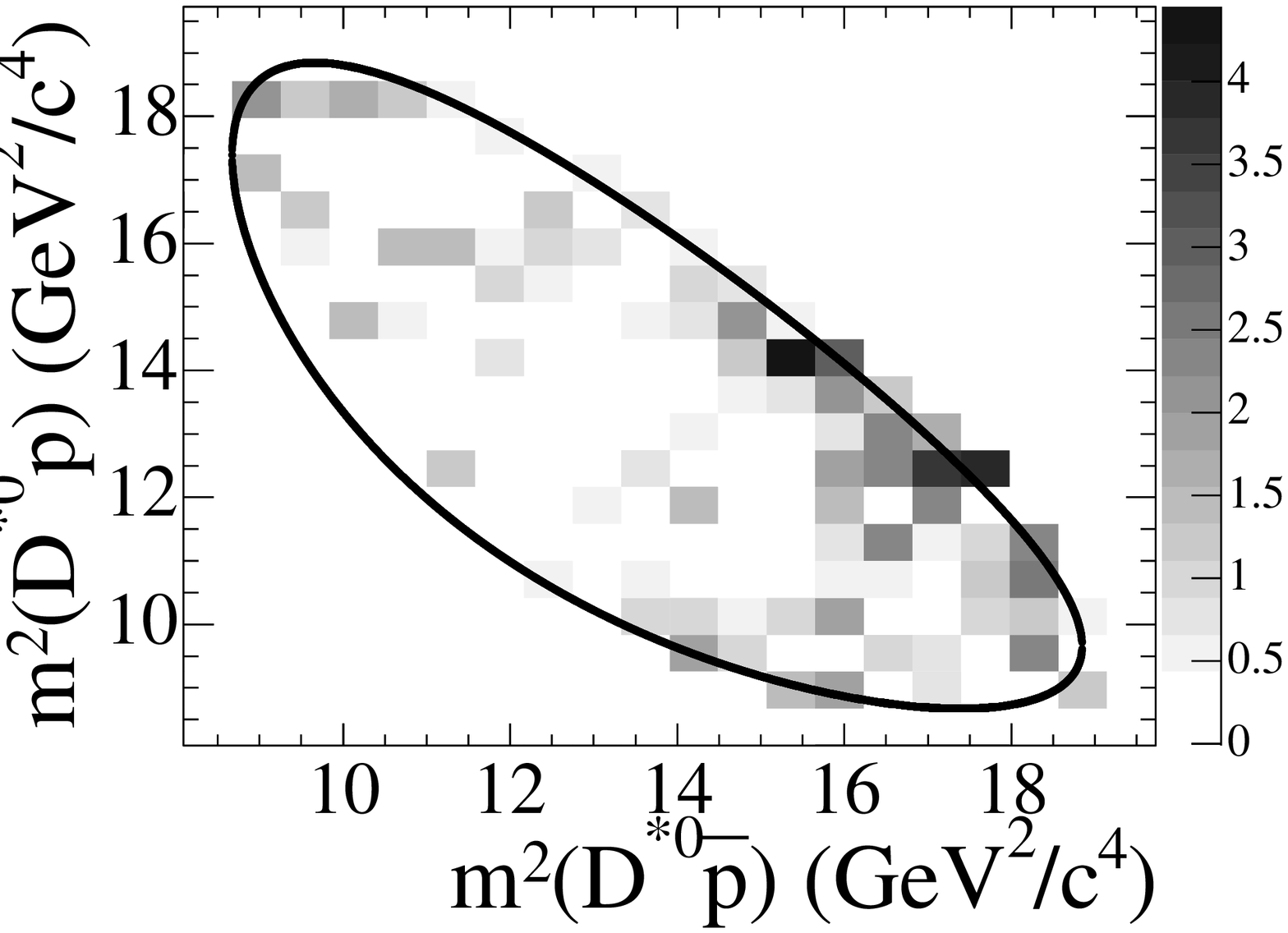}}
\centerline{\epsfxsize5.cm\epsffile{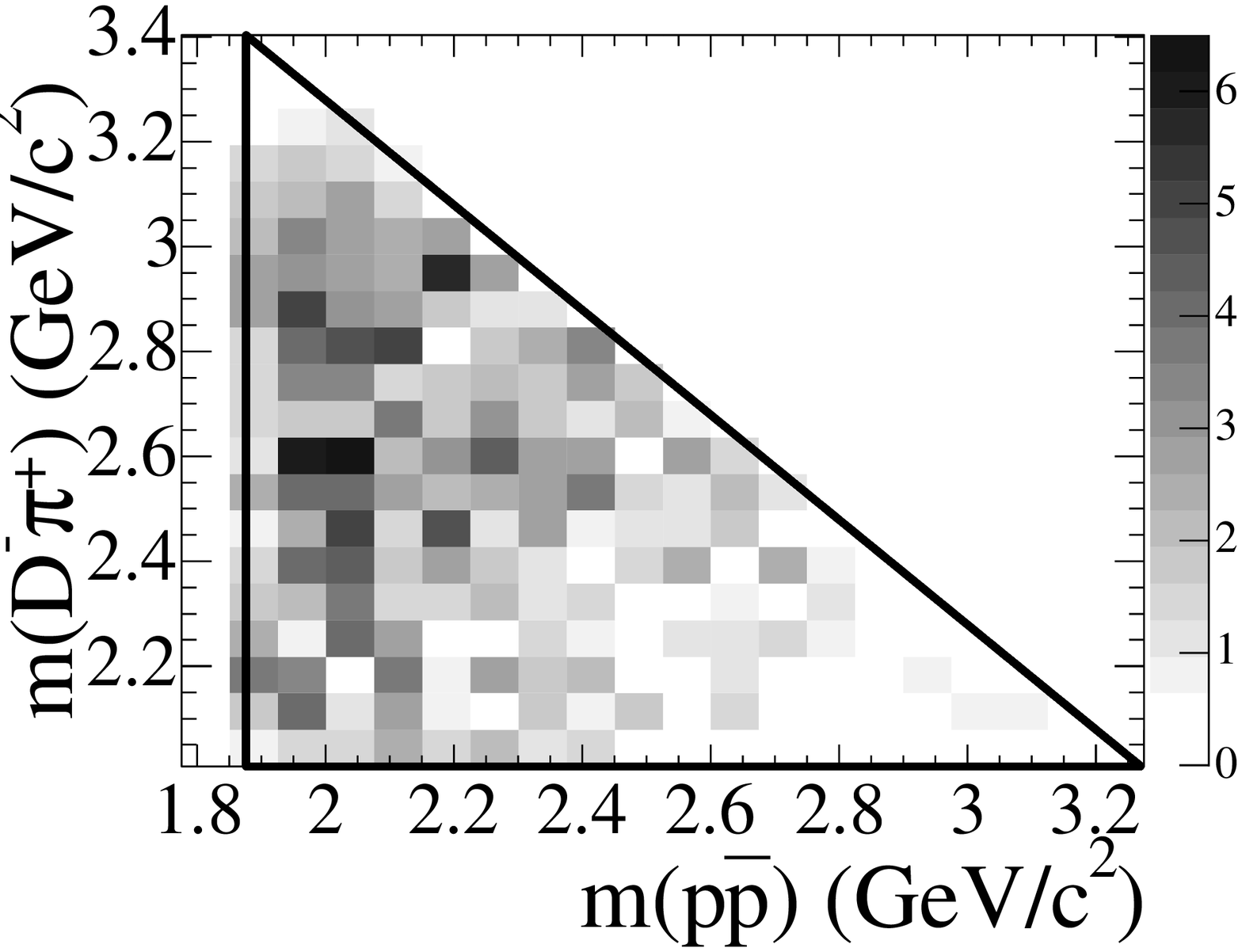}\hspace{0.1cm}
    \epsfxsize5.cm\epsffile{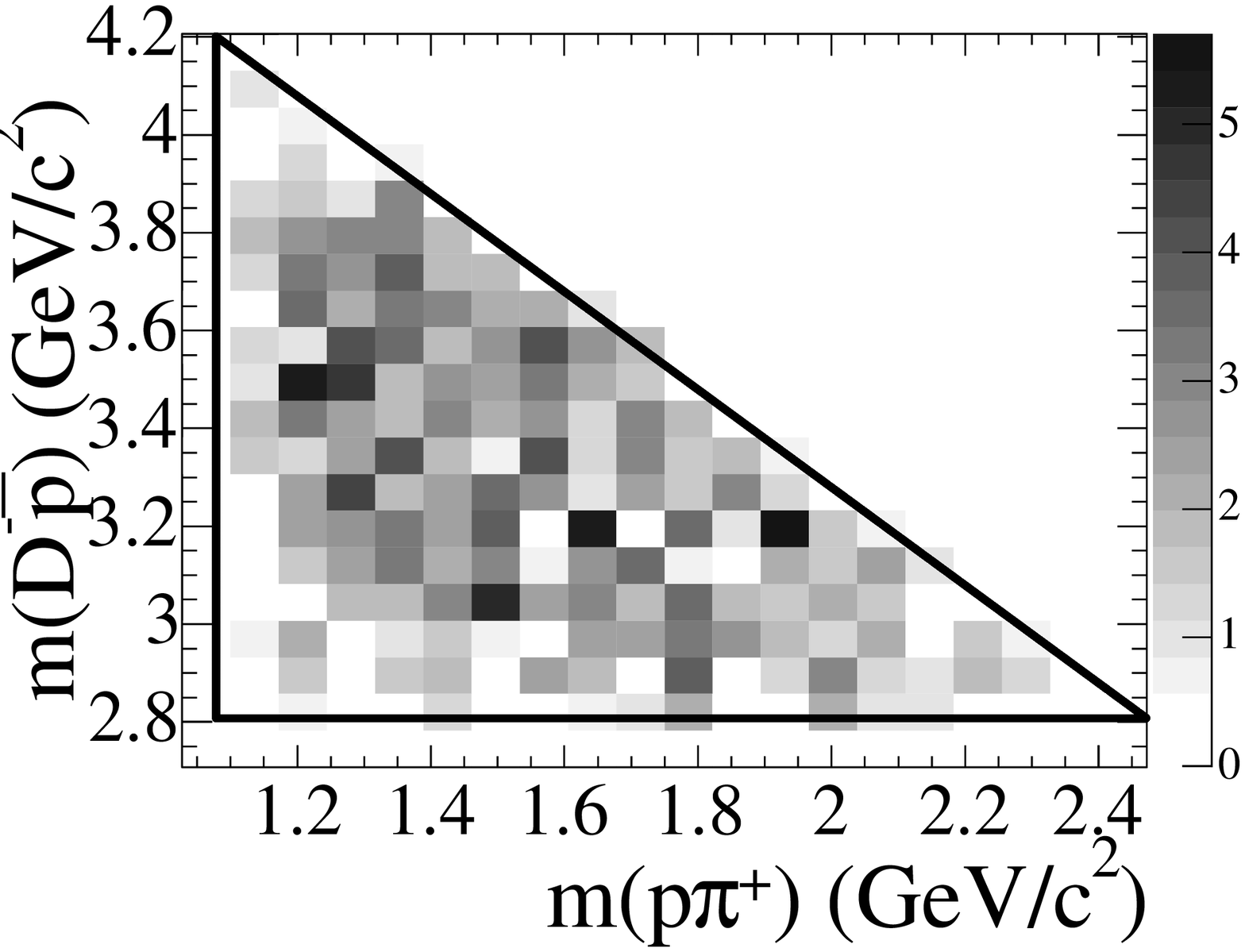}\hspace{0.1cm}
    \epsfxsize5.cm\epsffile{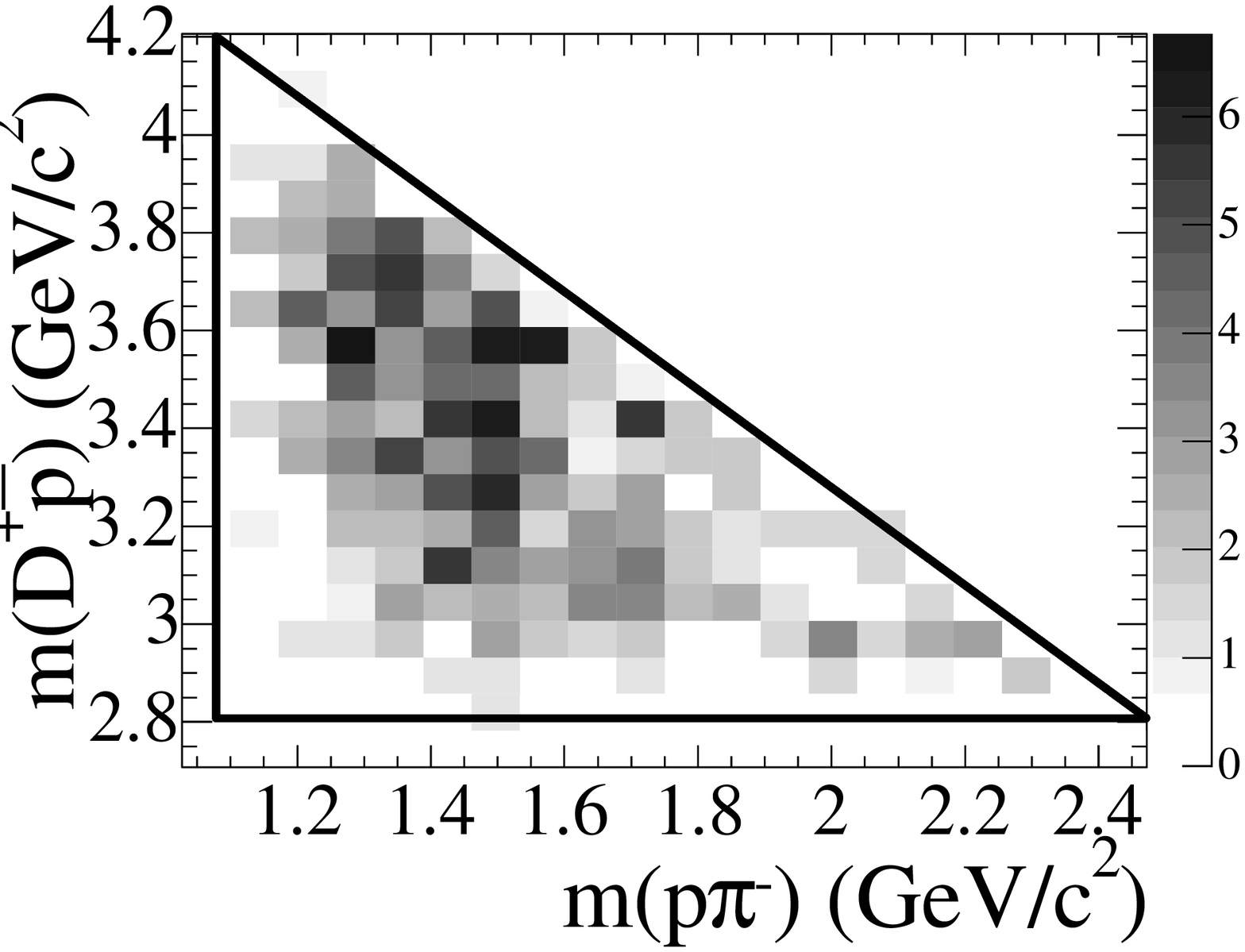}}
\centerline{\epsfxsize5.cm\epsffile{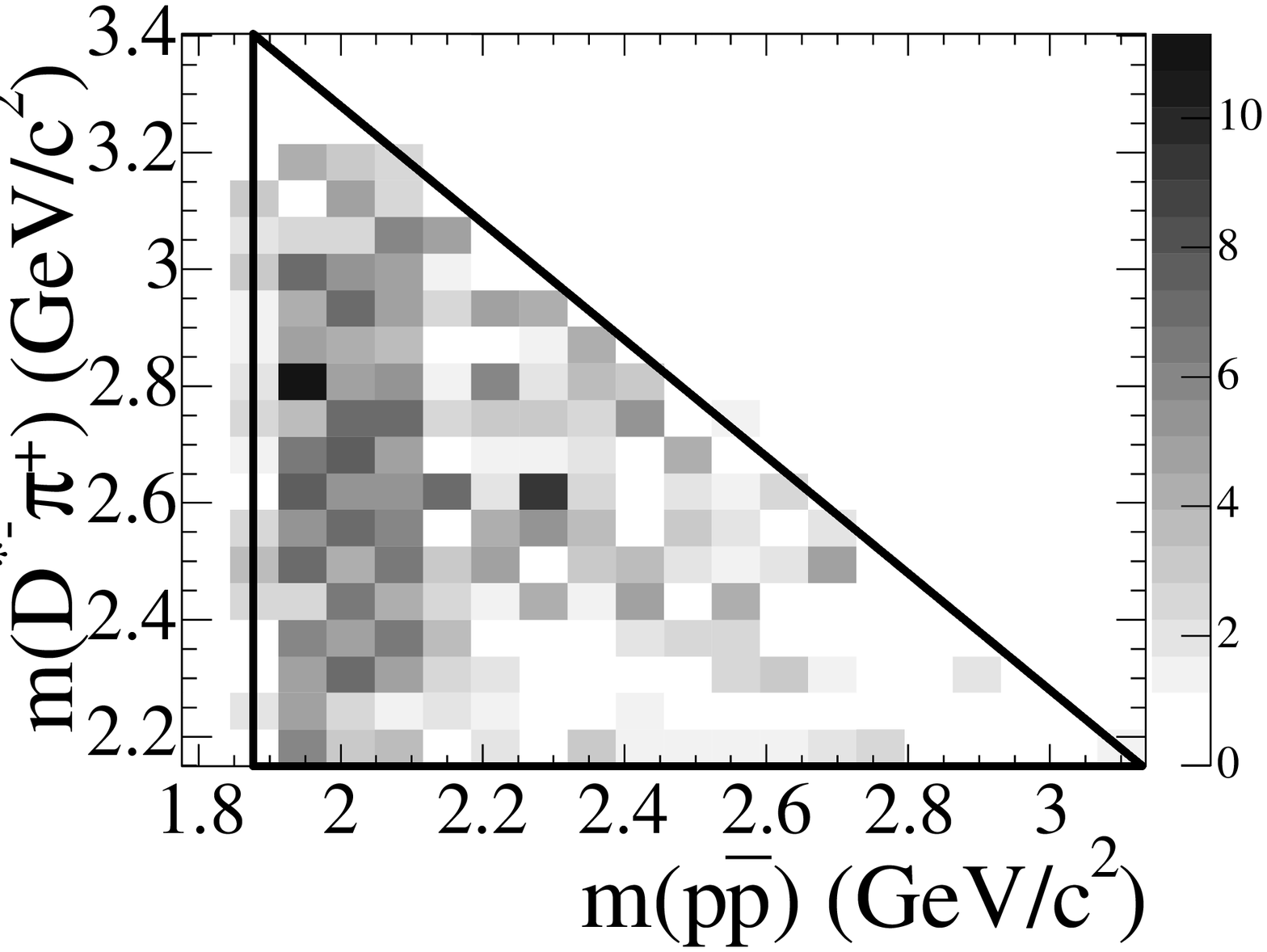}\hspace{0.1cm}
    \epsfxsize5.cm\epsffile{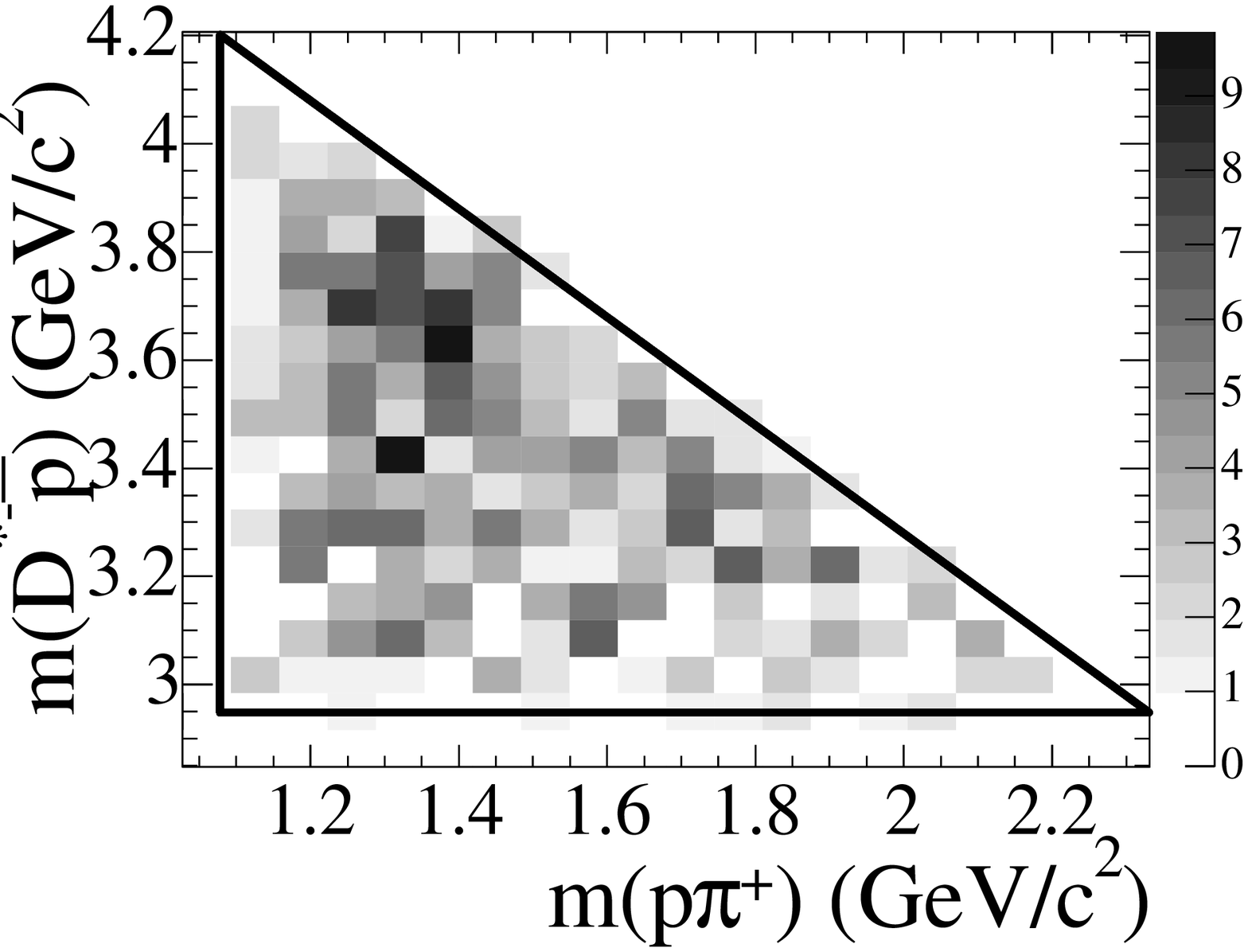}\hspace{0.1cm}
    \epsfxsize5.cm\epsffile{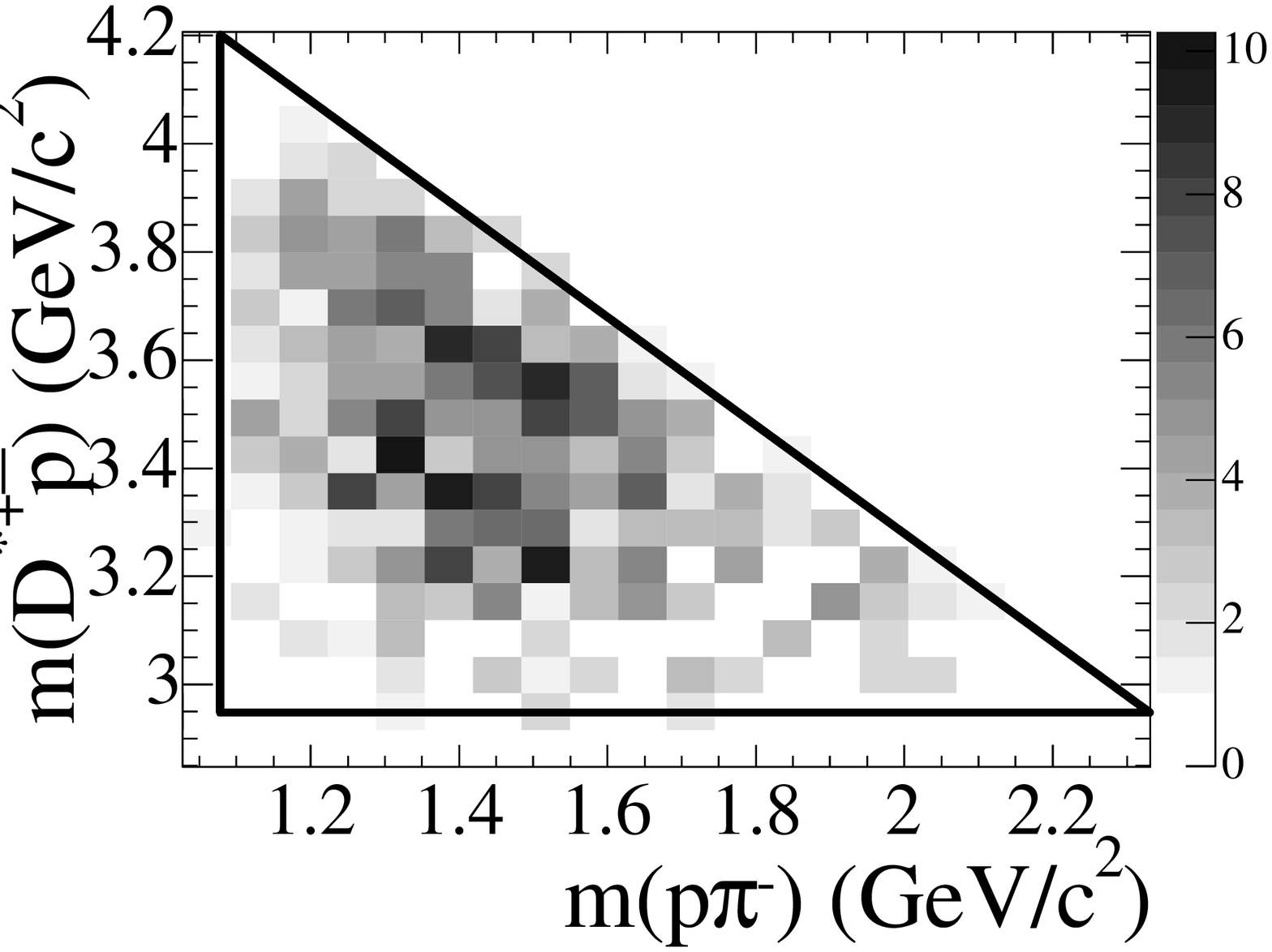}}
\end{figure*}

\begin{figure*}
 \caption{
\label{fig:allNt}The  branching fraction (B, in units of $10^{-6}/\gevcc$) distributions versus $p\overline{p}$ (top),   non-exotic ({\em i.e} $\overline{D}\overline{p}$) (middle), exotic ({\em i.e} $\overline{D}p$) (bottom) 
  invariant mass for
(from left) $\bdz$, $\bdsz$, $\bd$, and $\bds$  with all $D$ decay
modes combined.  The solid lines are the distributions expected from a purely phase-space decay. }
\centerline{\epsfxsize4.2cm\epsffile{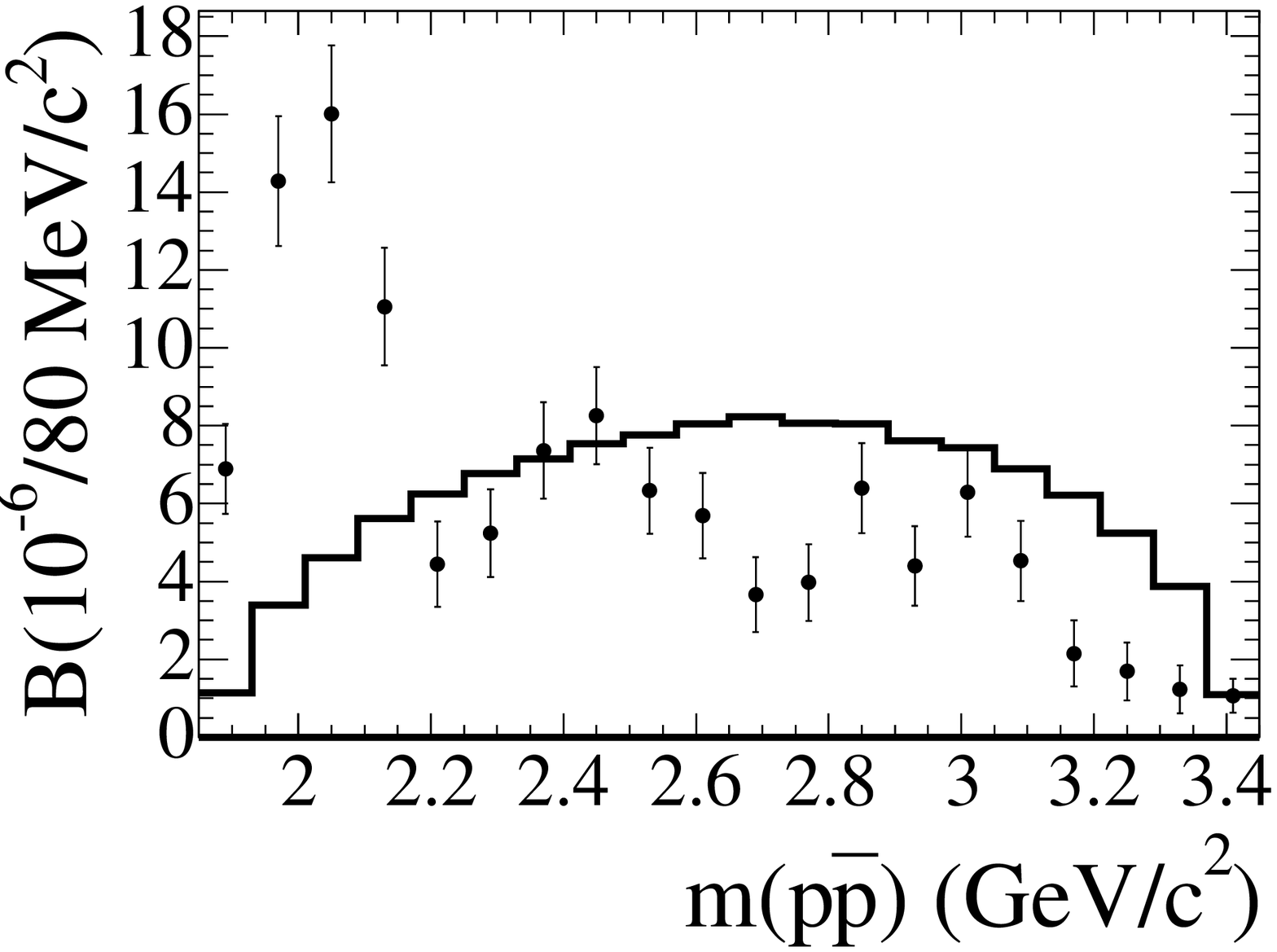}\hspace{0.1cm}
    \epsfxsize4.2cm\epsffile{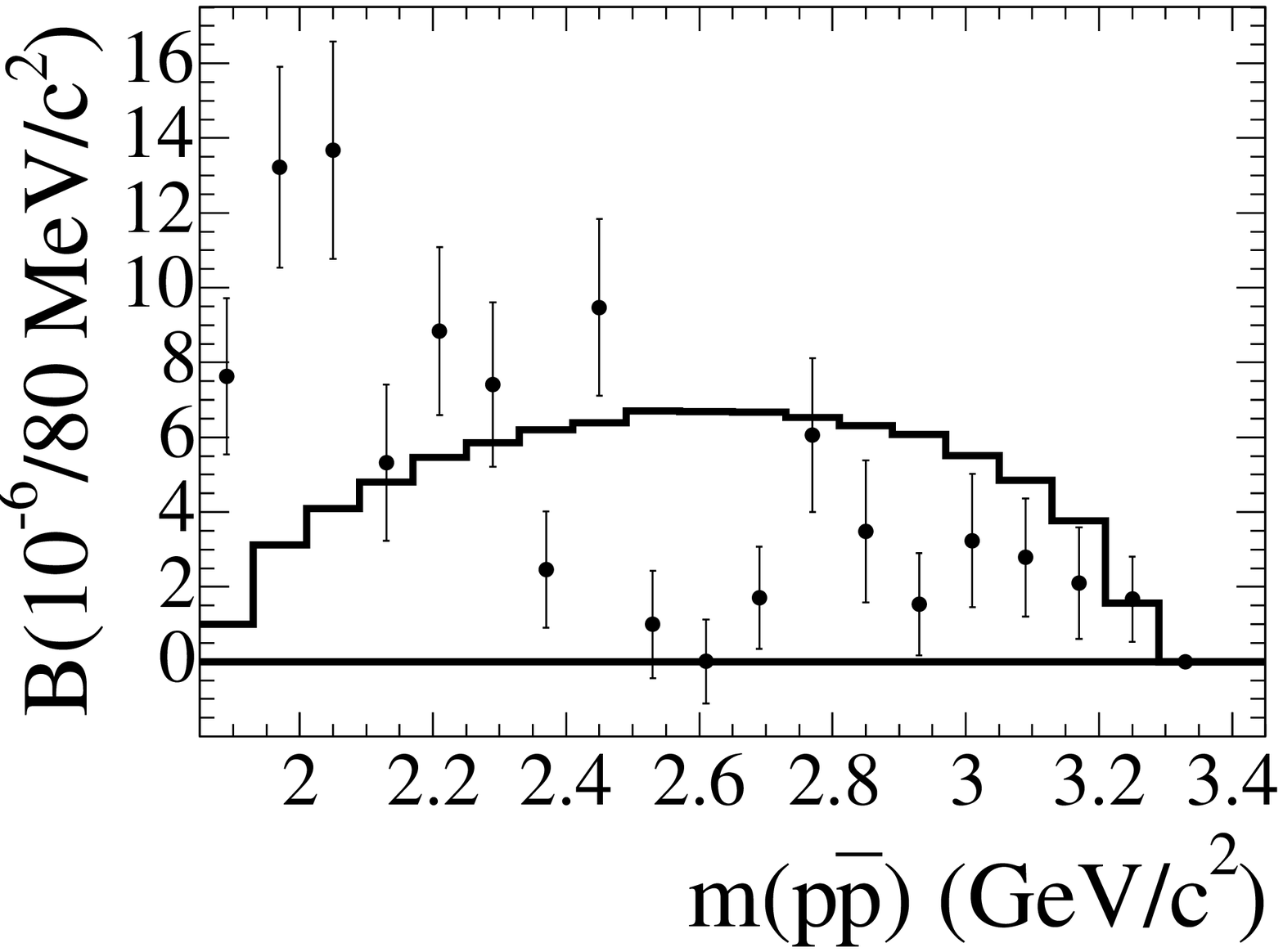}\hspace{0.1cm}
    \epsfxsize4.2cm\epsffile{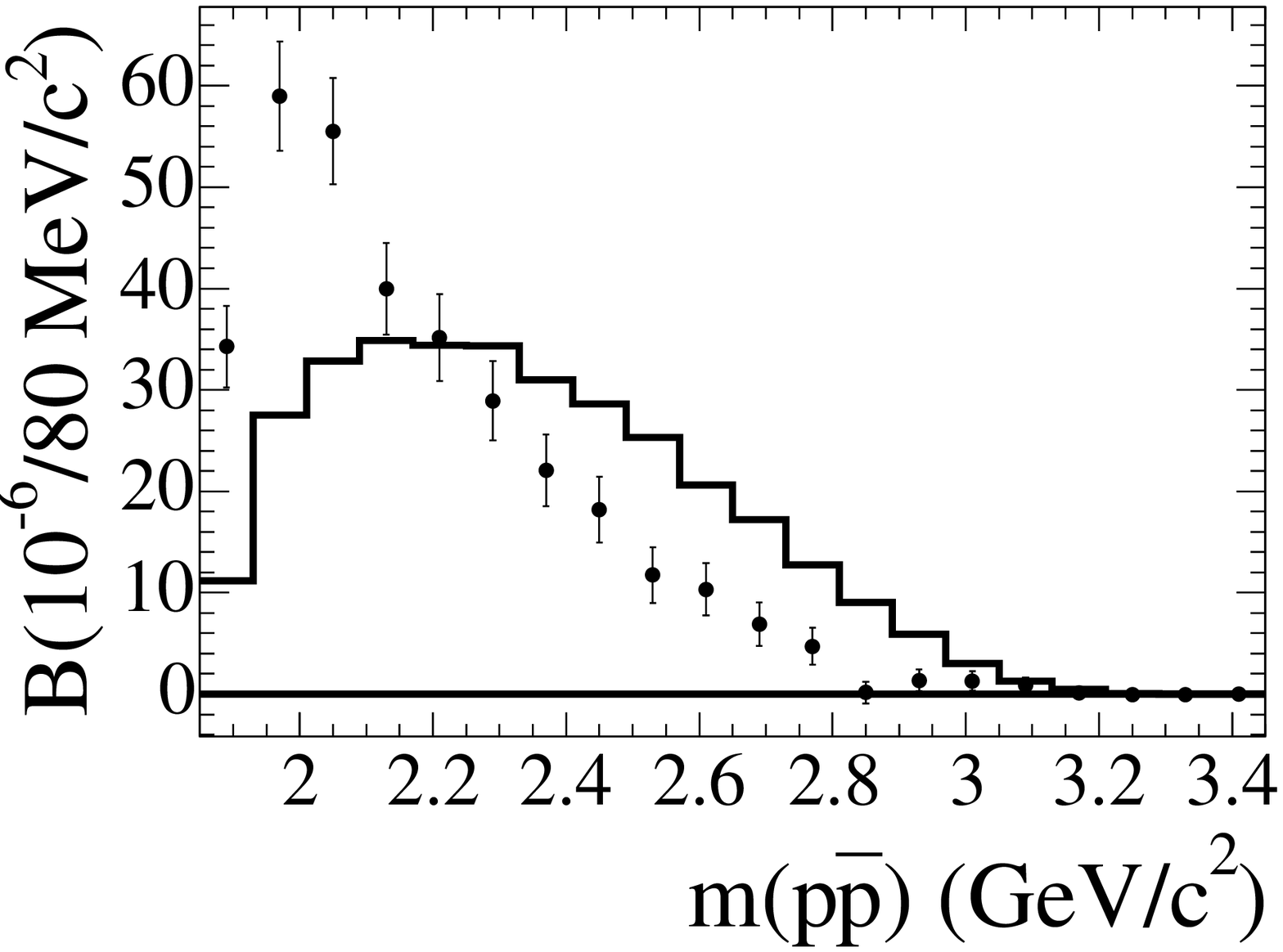}\hspace{0.1cm}
    \epsfxsize4.2cm\epsffile{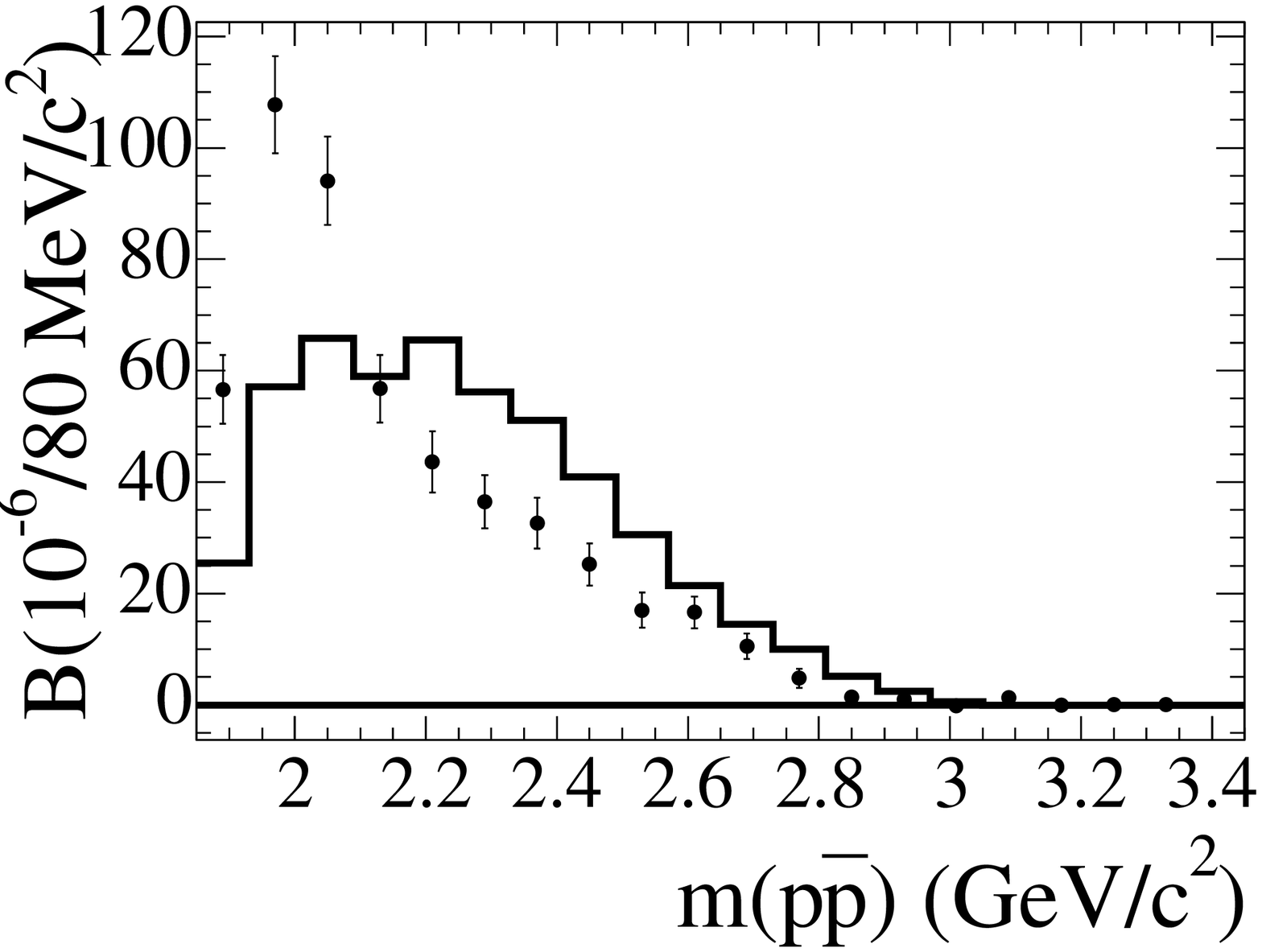}}
\centerline{\epsfxsize4.2cm\epsffile{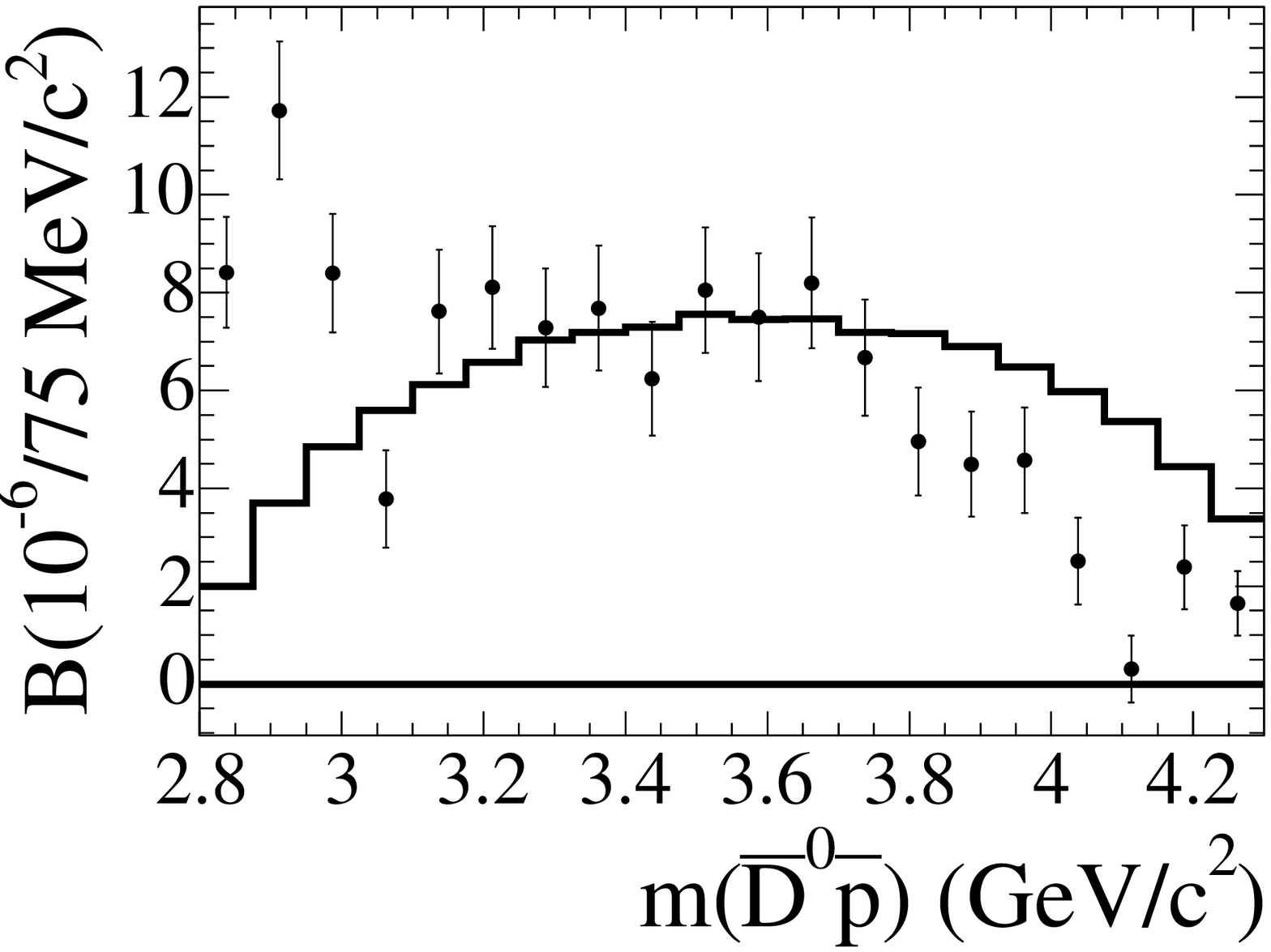}\hspace{0.1cm}
    \epsfxsize4.2cm\epsffile{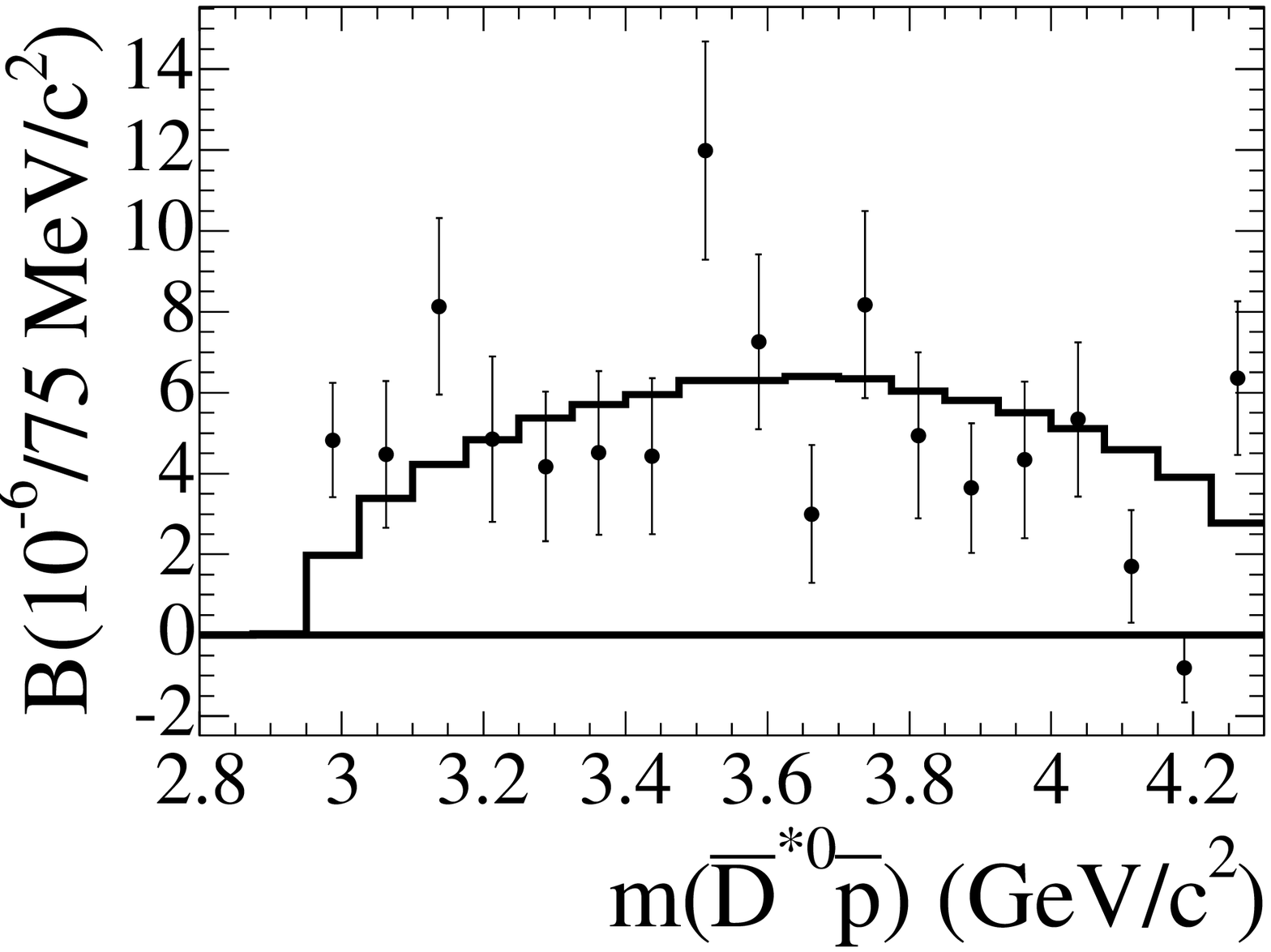}\hspace{0.1cm}
    \epsfxsize4.2cm\epsffile{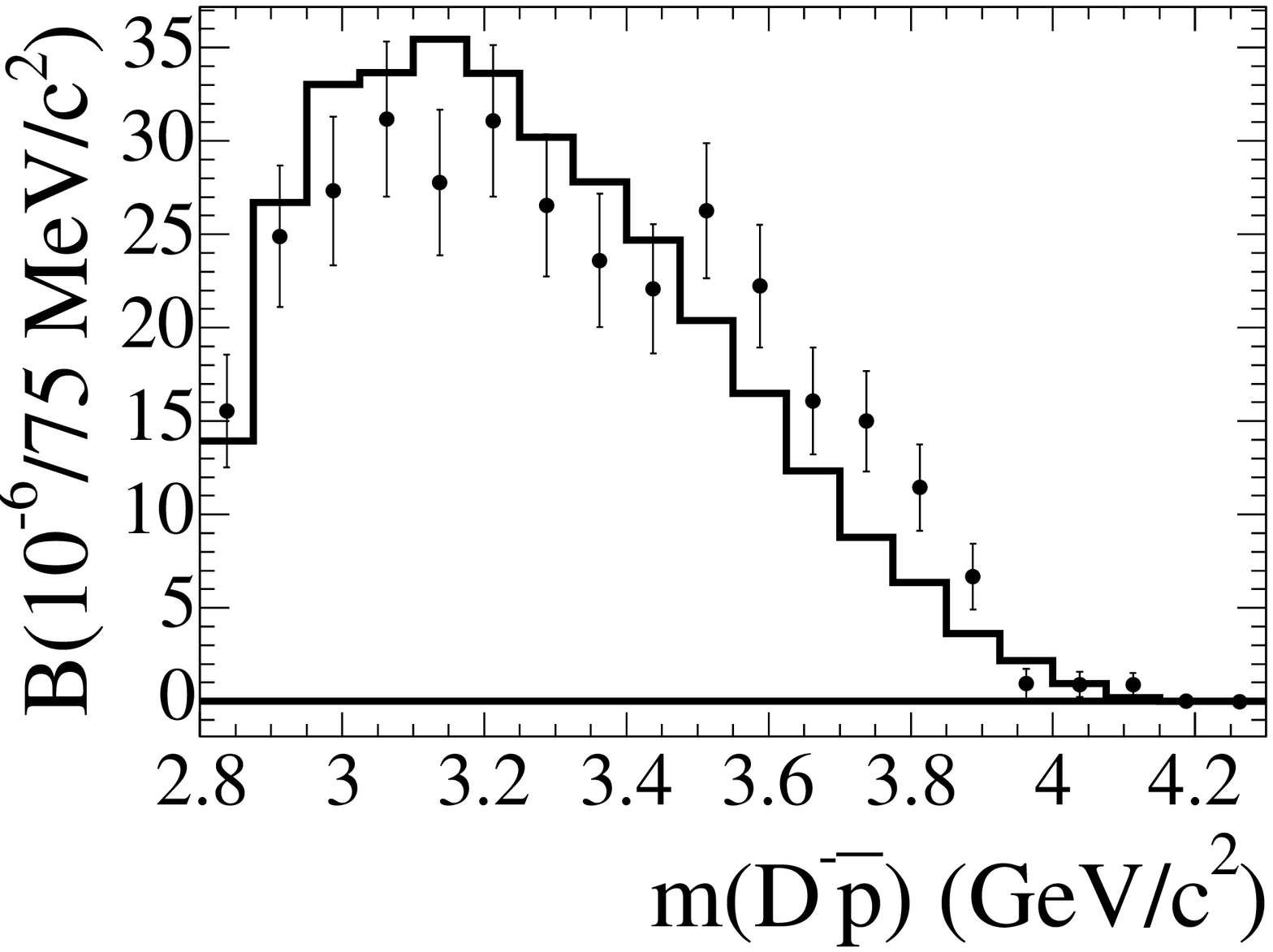}\hspace{0.1cm}
    \epsfxsize4.2cm\epsffile{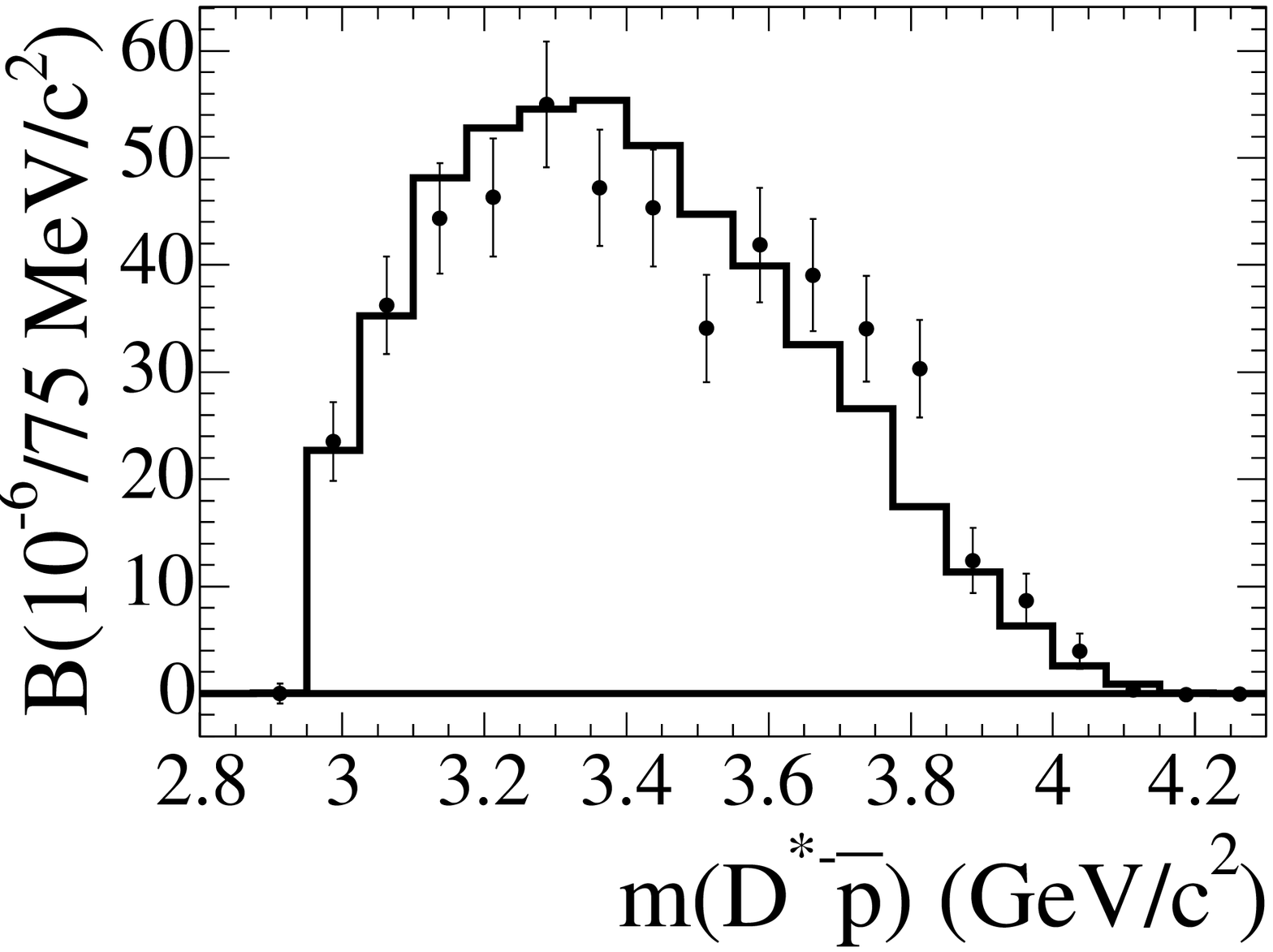}}
\centerline{\epsfxsize4.2cm\epsffile{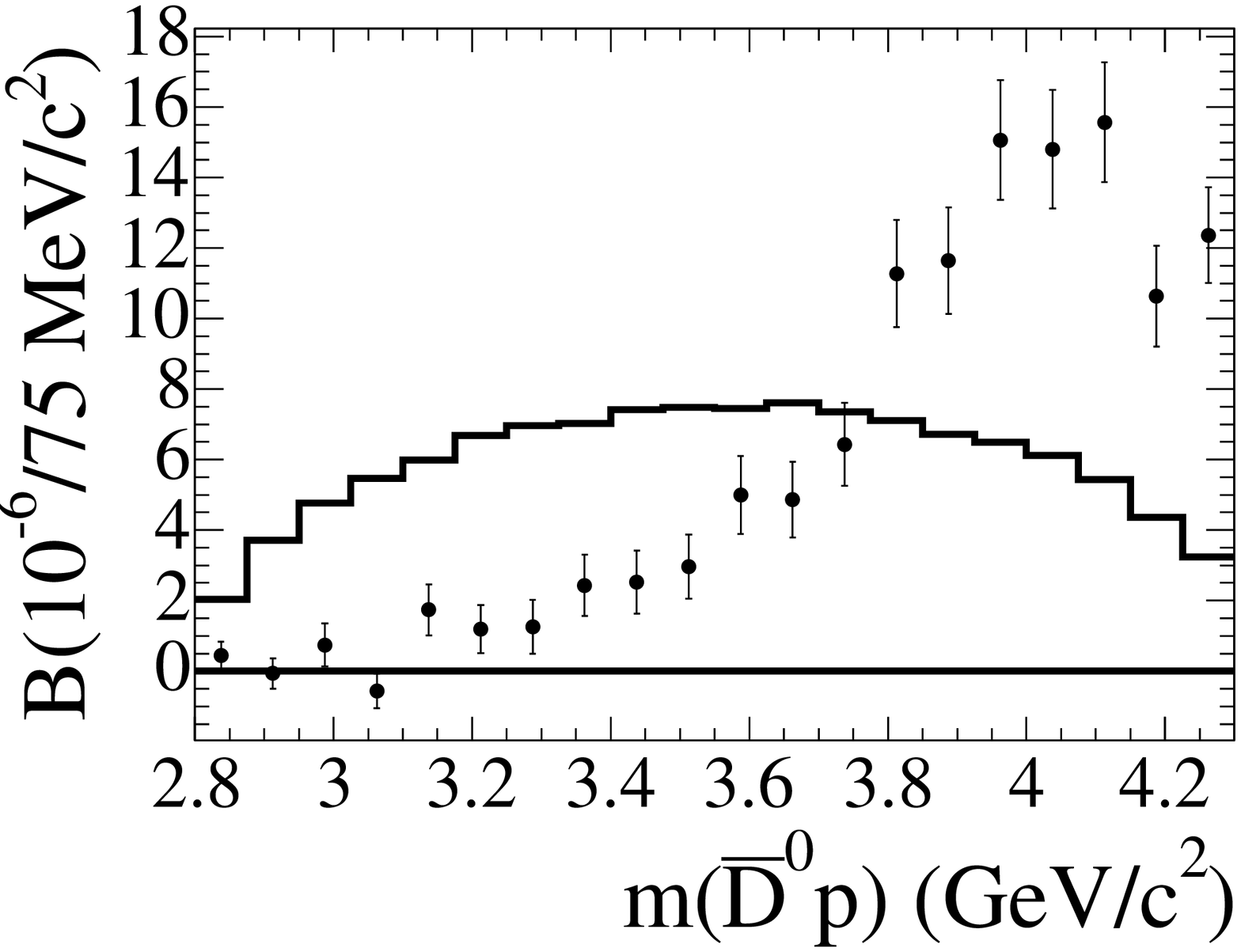}\hspace{0.1cm}
    \epsfxsize4.2cm\epsffile{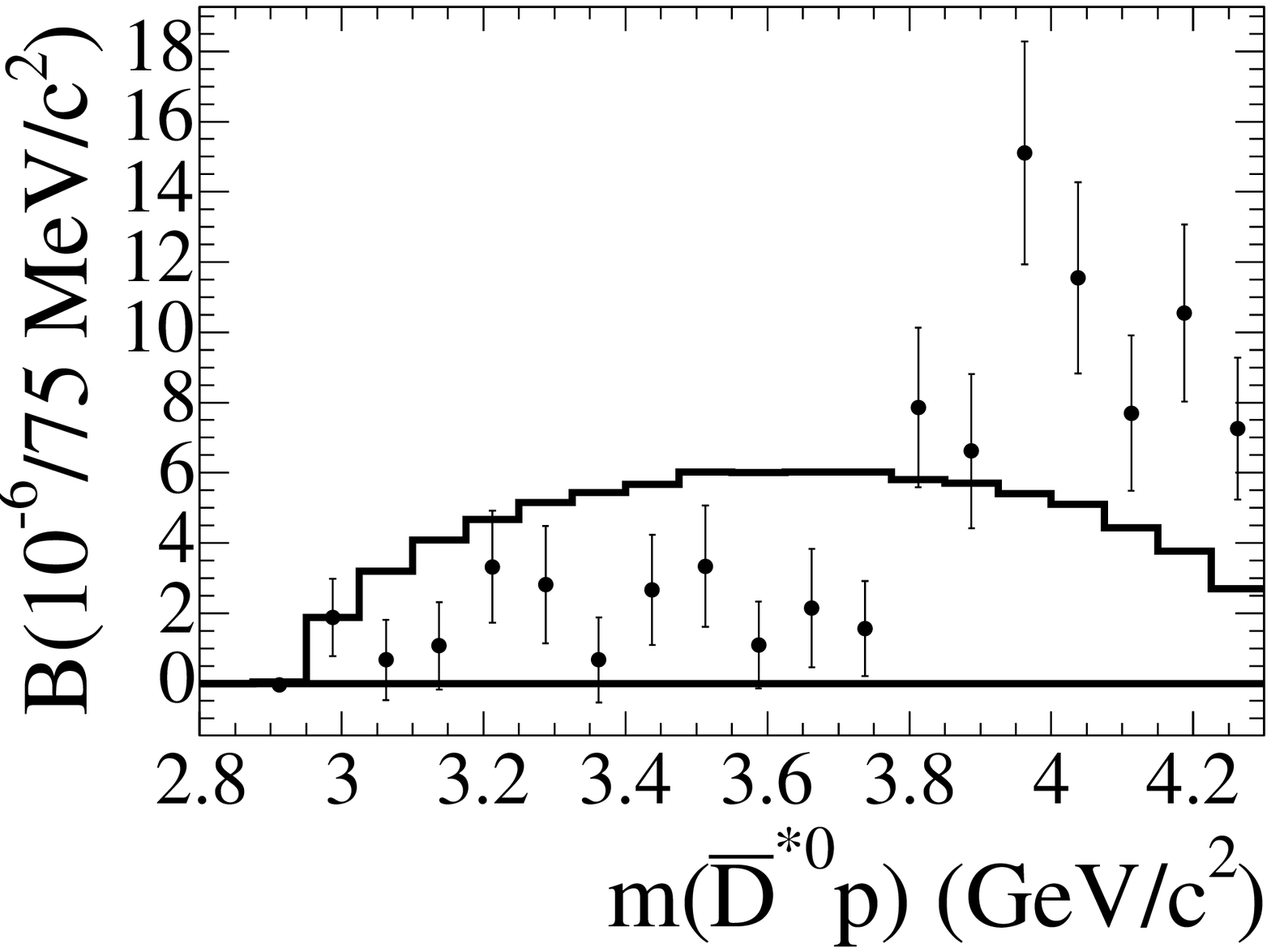}\hspace{0.1cm}
    \epsfxsize4.2cm\epsffile{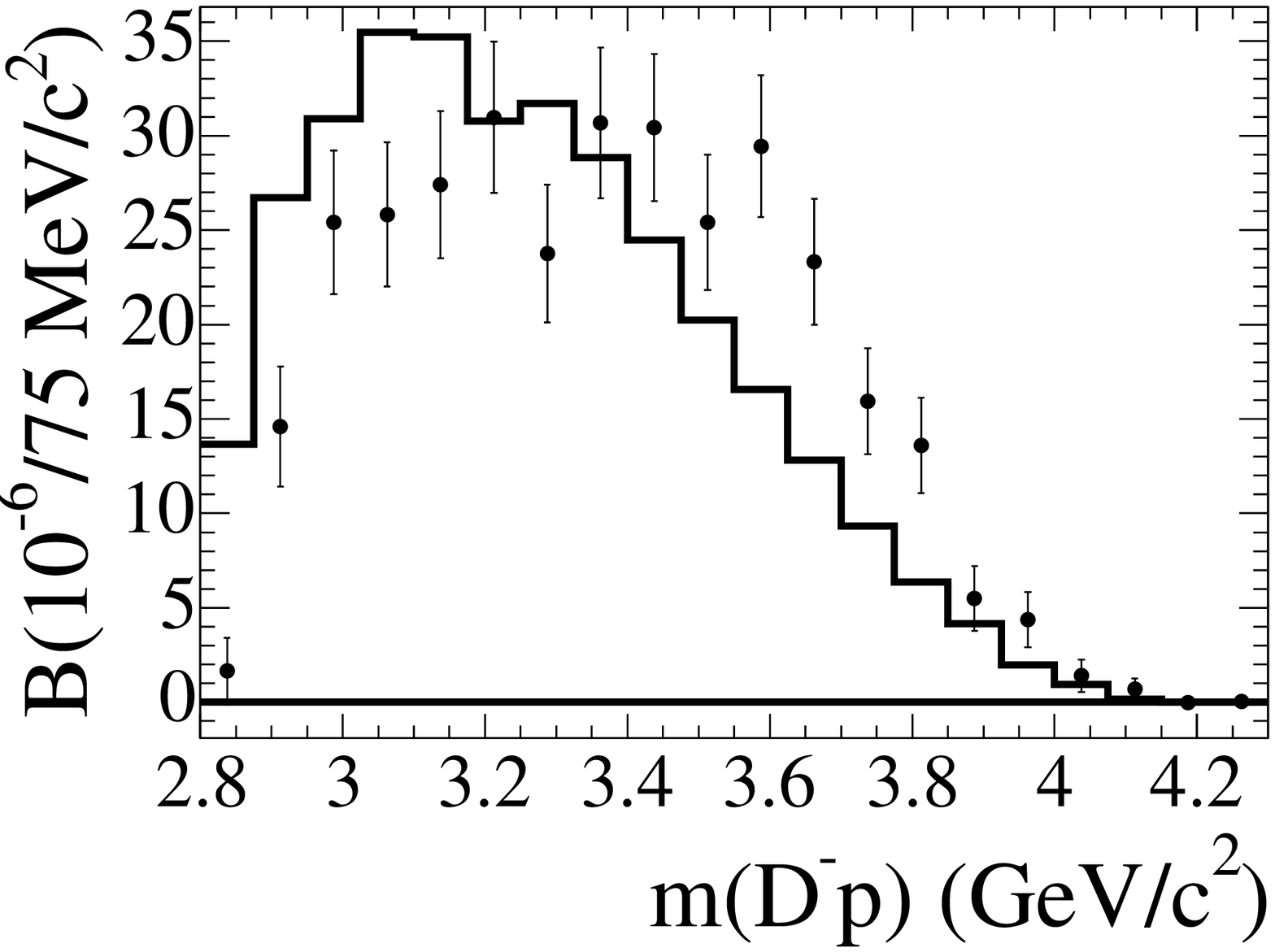}\hspace{0.1cm}
    \epsfxsize4.2cm\epsffile{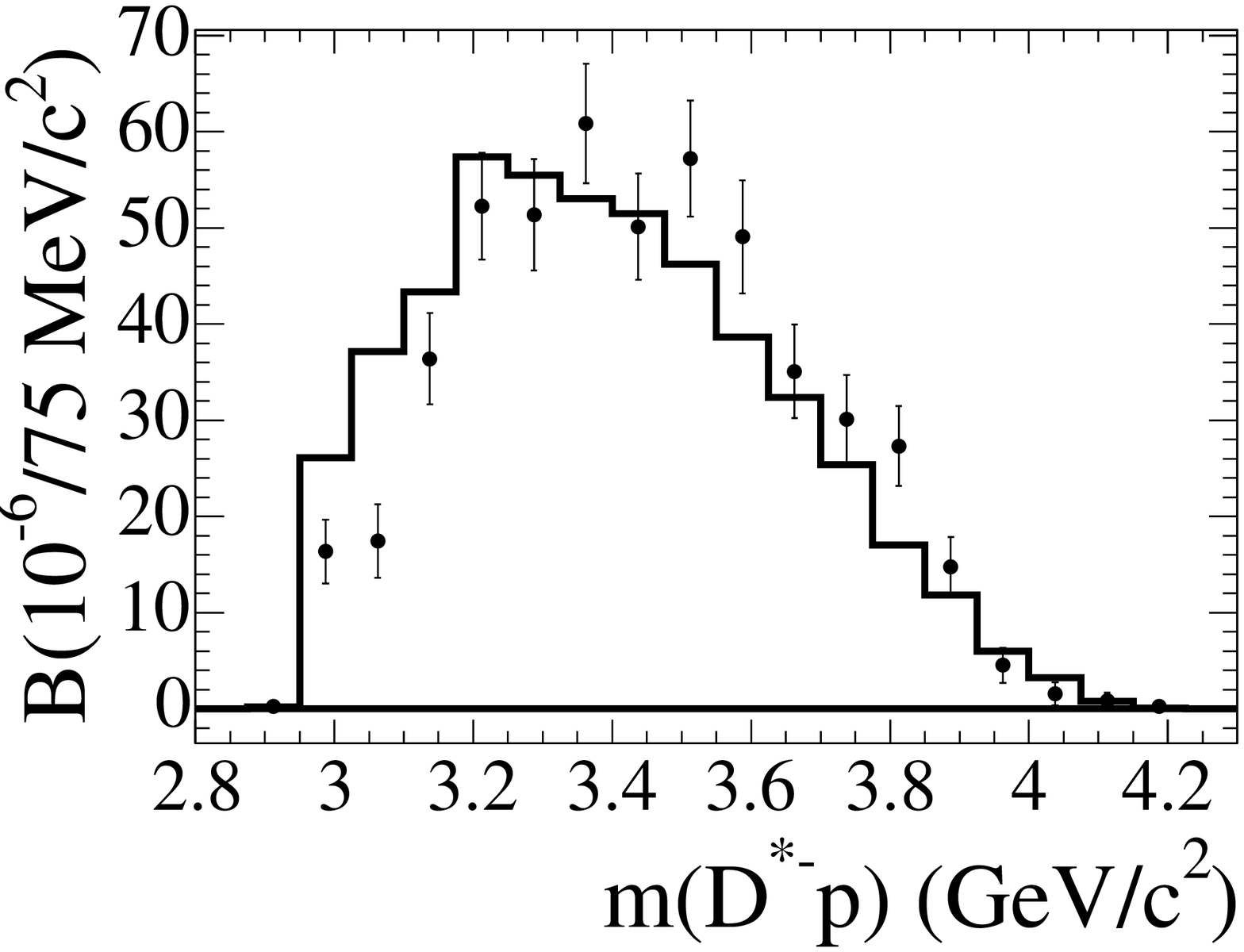}}
\end{figure*}

We investigate the decay dynamics by projecting the branching
fractions obtained with Equation \ref{eqn:br3} onto the different
invariant mass axes. This method requires that the variables used in the fit are uncorrelated to 
the variable being projected.  The correlations between the invariant masses and $\de$ and $\mes$ 
are observed to be small. 
Figure \ref{fig:dalitz} shows the two dimensional 
projections (the Dalitz plots for the 3-body decays) for the four modes under study.  
 Figure \ref{fig:allNt} shows 1-dimensional projections and the comparison with phase 
space distributions for the $p\overline{p}$, $\overline{D}\overline{p}$ (non-exotic minimal quark content of $\overline{c}\overline{u}\overline{d}$ or $\overline{c}\overline{d}\overline{d}$) and $\overline{D}p$ (exotic minimal quark content of $\overline{c}uuud$ or $\overline{c}duud$) invariant masses. 

\begin{figure*}
 \caption{
\label{fig:ComparePP}Left:  The phase space-corrected $p\overline{p}$ invariant mass distributions for all four decay modes:  $\bdz$ (triangles), $\bdsz$ (open circles), $\bd$ (squares), and $\bds$ (closed circles).  Right:   The  $p\overline{p}$ distributions from the present analysis averaged over the four decay modes (closed circles) compared to the distributions obtained in $e^+e^-\rar p\overline{p}\gamma$ (open squares) and $B^+\rar p\overline{p}K^+$ (open circles).  
}
    \epsfxsize8.cm\epsffile{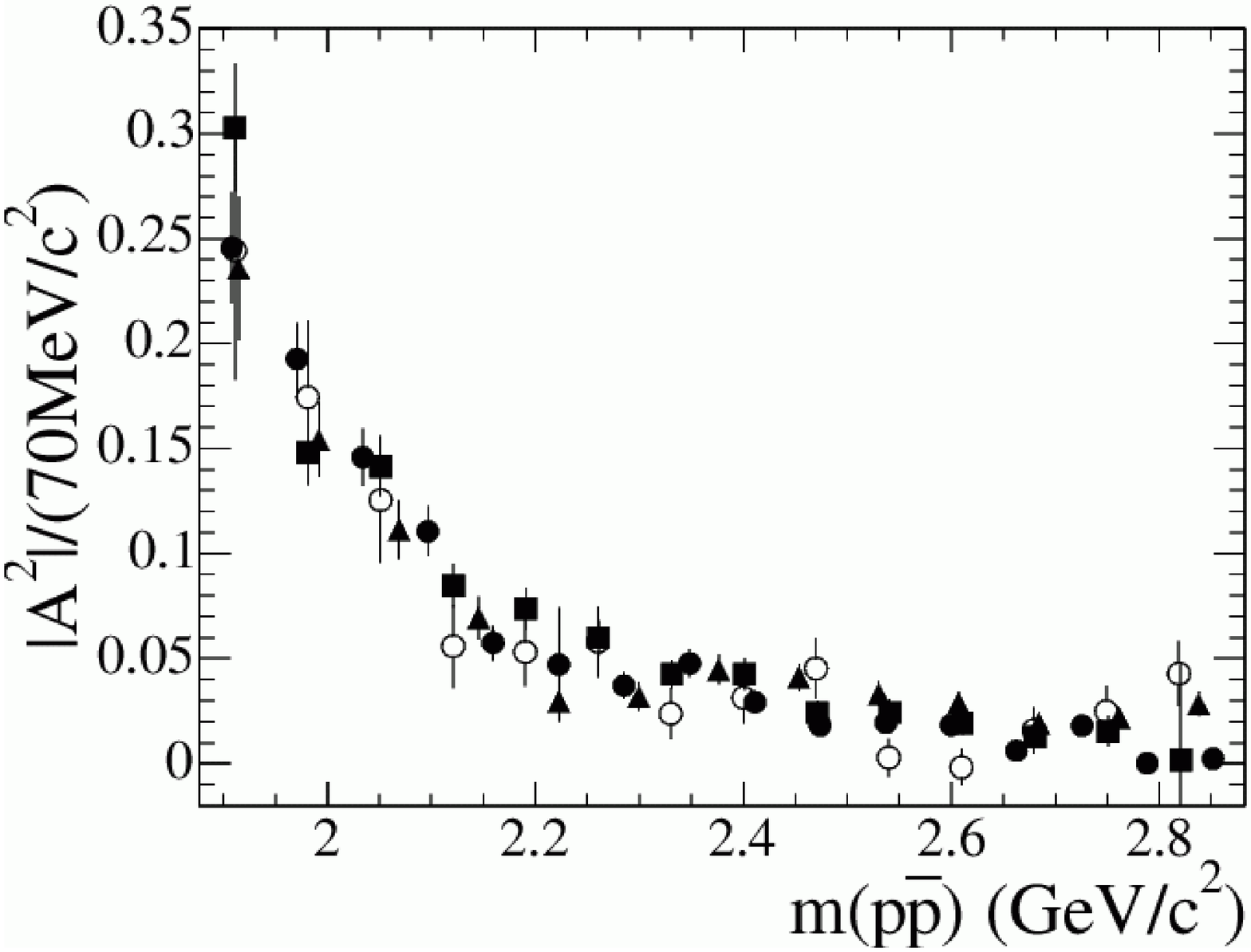}
    \epsfxsize8.cm\epsffile{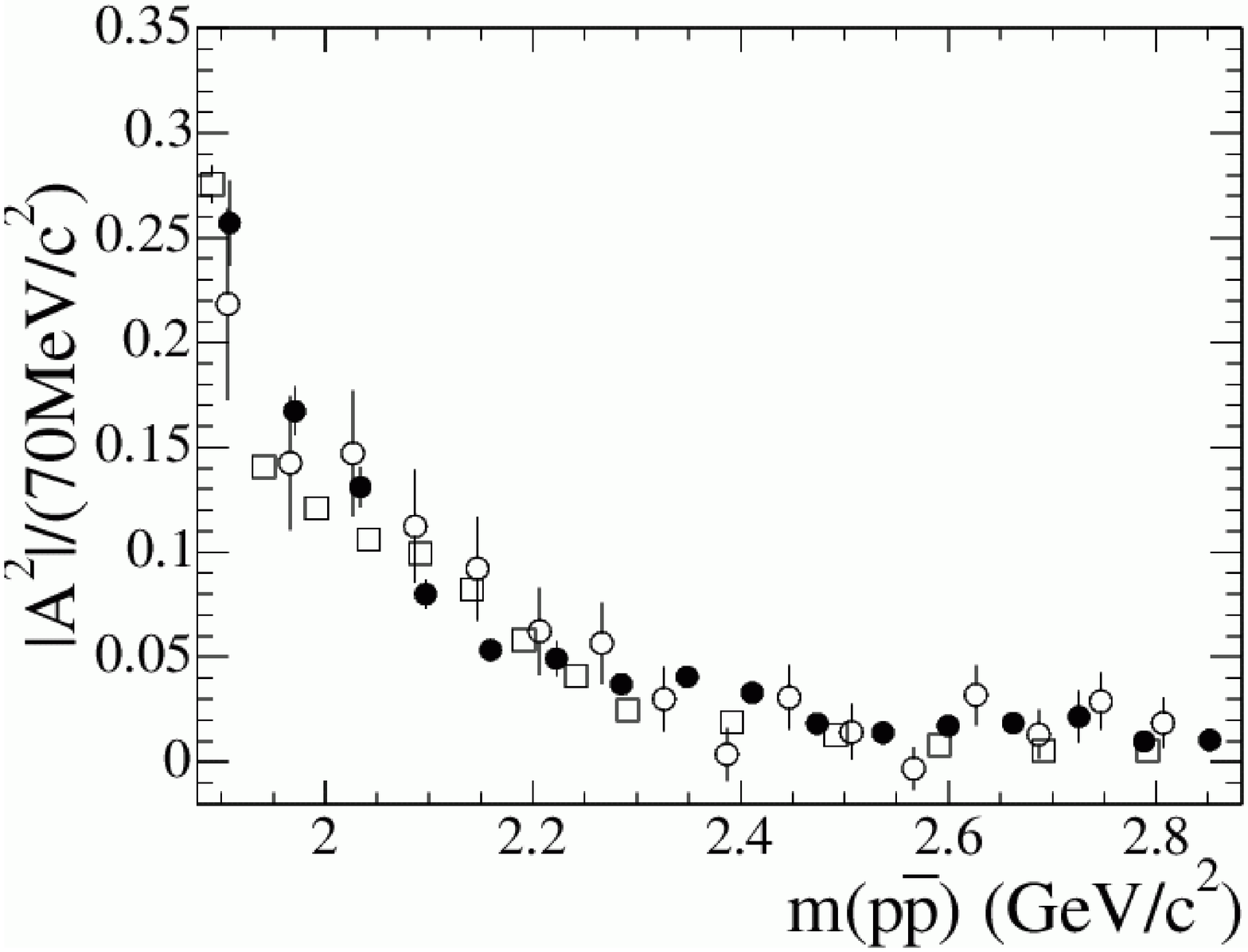}

\end{figure*}

In comparison with phase space, an enhancement at low $p\overline{p}$ mass is seen in all decay channels. 
Such an enhancement has been observed in other situations \cite{Abe:2002ds,Aubert:2005gw,Aubert:2005cb,prl_91_022001};
 indeed, it is also observed in the background $p\overline{p}$ distributions in this analysis.  
In the left plot of Figure \ref{fig:ComparePP} the 
$p\overline{p}$  distributions for all four modes have been overlaid removing the
events with $M(\overline{D}^{(*)}\overline{p})$ less than 3.1 \gevcc and normalizing
to the total area.  In addition, each event entering  Figure \ref{fig:ComparePP} has been weighted by a phase-space 
factor and  thus the distribution is proportional to the square of the matrix element. 
The distributions of the four modes show the same behavior.  
We have also compared our phase-space corrected $p\overline{p}$ distributions (averaged over the four modes) 
to those measured in $e^+e^-\rar p\overline{p}\gamma$ \cite{Aubert:2005cb} and $B^+\rar p\overline{p}K^+$ \cite{Aubert:2005gw} by \babar, shown on the right of  Figure \ref{fig:ComparePP}, and again there appears to be
good agreement. 

Explanations that have been proposed to account for the enhancement observed at the $p\overline{p}$ threshold include a gluonic resonance \cite{chua} and short-range correlations between the $p$ and the $\overline{p}$ \cite{rosner2}.  The BES collaboration  has recently claimed  evidence for a resonance 
decaying to $\pi\pi\eta^\prime$ with a mass of 1834 \mevcc and a width of 69 \mevcc \cite{Ablikim:2005um}. 
This resonance should also decay to $p\overline{p}$ and 
the mass and width measured by BES in  $\pi\pi\eta^\prime$ is in  agreement with the 
enhancement seen by BES in the $p\overline{p}$ distribution in J$/\psi\to\gamma p \overline{p}$ decays \cite{prl_91_022001} assuming a Breit-Wigner with corrections for final state interactions \cite{zou,ulf}.   

With respect to the $\overline{D}\overline{p}$ invariant mass spectra, other than an excess at low mass in the $\bdz$ mode, the  plots in the middle row of Figure \ref{fig:allNt} are in qualitative agreement with the phase space histograms. The low mass excess in $\bdz$ is also easily seen in the Dalitz plot in Figure \ref{fig:dalitz} and appears again to be a threshold enhancement as in the $p\overline{p}$ case.  
While it would be expected that the same effect would be seen in the $\bdsz$ mode, the statistics are much lower and the mass threshold is higher.  

The $\overline{D}p$ distributions, in the bottom row of Figure \ref{fig:allNt}, we observe a  clear tendency to peak toward high $\overline{D}^{(*)0}p$ mass in comparison with phase space for the three-body modes.  This is also reflected in the apparent asymmetry in the Dalitz plots.  The four body modes are in qualitative agreement with phase space distributions in the $\overline{D}p$ projections.  

The H1 Collaboration has claimed evidence for a charmed pentaquark
state decaying to $D^{*-}p$ at 3.1 \gevcc whose width is  less than
their experimental resolution of 7.1 \mevcc. By fitting the $D^{-}p$
invariant mass spectrum in the decay $\bd$ to a Breit-Wigner plus linear background, 
we obtain an upper limit on the branching fraction:
\begin{equation}
{\cal B} (\Bz \to \Theta_c \antiproton\pi^+)\times{\cal B}(\Theta_c \to D^-\proton)<9\times10^{-6},
\end{equation}
while for the  $D^{*-}p$ spectrum in $\bds$ we obtain:
\begin{equation}
{\cal B} (\Bz \to \Theta_c \antiproton\pi^+)\times{\cal B}(\Theta_c  \to \Dstarm\proton)<14\times10^{-6}
\end{equation}
at 90\% C.L. For this limit we have assumed the resonance 
width for the $\Theta_c$ to be 25 \mevcc, which corresponds to the upper limit
on the width given by H1. If we assume a smaller width, the limits decrease. 

In conclusion, we have measured the branching fractions of $\bdz$,
$\bdsz$, $\bd$, and $\bds$. The results obtained for the modes
$\bds$, $\bdsz$, and $\bdz$ agree with the previous measurements and have 
smaller uncertainties
while the decay $\bd$ has been measured for the first time. We do
not observe any evidence for the charmed pentaquark observed by H1
at $M(D^{*-}p)$ of 3.1 \gevcc. In comparison with phase space 
we observe a low-mass $p\overline{p}$ enhancement  similar to 
other observations in $p\overline{p}$ production.  We also observe a deviation from phase-space
structure in the $Dp$ and $\overline{D}p$ invariant mass distributions for the three-body modes. 

We are grateful for the excellent luminosity and machine conditions
provided by our \pep2\ colleagues, 
and for the substantial dedicated effort from
the computing organizations that support \babar.
The collaborating institutions wish to thank 
SLAC for its support and kind hospitality. 
This work is supported by
DOE
and NSF (USA),
NSERC (Canada),
IHEP (China),
CEA and
CNRS-IN2P3
(France),
BMBF and DFG
(Germany),
INFN (Italy),
FOM (The Netherlands),
NFR (Norway),
MIST (Russia), and
PPARC (United Kingdom). 
Individuals have received support from CONACyT (Mexico), A.~P.~Sloan Foundation, 
Research Corporation,
and Alexander von Humboldt Foundation.


\begin{thebibliography}{99}
\bibitem{chconj} Throughout this paper, the named reaction refers also to its charge conjugate.
\end{thebibliography}
\end{document}